\documentclass[final,5p,times,twocolumn]{elsarticle}

\usepackage{graphicx}
\usepackage{subfigure}
\usepackage{multirow}
\usepackage{wrapfig}

\usepackage{tabularray}
\usepackage{tabularx}
\usepackage{rotating}
\usepackage{makecell}
\usepackage{booktabs}

\usepackage[utf8]{inputenc}
\usepackage{amsmath}
\usepackage{amsthm}
\usepackage{amsfonts}
\usepackage{epsfig}
\usepackage{psfrag}
\usepackage{pstricks}
\usepackage{algorithm}

\usepackage{algorithmicx}
\usepackage{algpseudocode}  
\floatname{algorithm}{Algorithm}

\graphicspath{{./Figures/}}

\usepackage{amssymb}
\usepackage{bm}
\biboptions{comma,sort&compress}

\usepackage{enumitem}
\setlist{nosep,topsep=-\parskip}

\usepackage{color}
\usepackage{ulem}
\graphicspath{{figures/},{pics/}}
\newcommand{\rev}[2]{#2}

\journal{CAD}

\begin{document}

\begin{frontmatter}

\title{Warp-centric GPU meta-meshing and fast triangulation of billion-scale lattice structures}
\author[]{Qiang Zou\corref{cor}}\ead{qiangzou@cad.zju.edu.cn}
\author[]{Yunzhu Gao}

\cortext[cor]{Corresponding author.}
\address{State Key Laboratory of CAD\&CG, Zhejiang University, Hangzhou, 310027, China}

\begin{abstract}
Lattice structures have been widely used in applications due to their superior mechanical properties. To fabricate such structures, a geometric processing step called triangulation is often employed to transform them into the STL format before sending them to 3D printers. Because lattice structures tend to have high geometric complexity, this step usually generates a large amount of triangles, a memory and compute-intensive task. This problem manifests itself clearly through large-scale lattice structures that have millions or billions of struts. To address this problem, this paper proposes to transform a lattice structure into an intermediate model called meta-mesh before undergoing real triangulation. Compared to triangular meshes, meta-meshes are very lightweight and much less compute-demanding. The meta-mesh can also work as a base mesh reusable for conveniently and efficiently triangulating lattice structures with arbitrary resolutions. A CPU+GPU asynchronous meta-meshing pipeline has been developed to efficiently generate meta-meshes from lattice structures. It shifts from the thread-centric GPU algorithm design paradigm commonly used in CAD to the recent warp-centric design paradigm to achieve high performance. This is achieved by a new data compression method, a GPU cache-aware data structure, and a workload-balanced scheduling method that can significantly reduce memory divergence and branch divergence. Experimenting with various billion-scale lattice structures, the proposed method is seen to be two orders of magnitude faster than previously achievable.
\end{abstract}

\begin{keyword} 
CAD; Geometric modeling; Lattice structures; Meta-meshing; Triangulation; GPU computing 
\end{keyword}

\end{frontmatter}


\section{Introduction}
\label{sec:intro}
With the advances in additive manufacturing, lattice structures have seen increasing applications in fields like automotive, shipbuilding, and aerospace industries~\cite{2021_Liu_Memory-efficient,2021_Ding_STL-free}. The network of interconnected struts in lattice structures offers huge potential for design optimization, e.g., lightweight, high strength, multifunctional~\cite{2022_Liu_lattice_aerospace,2012_lattice-heat-exchanger,2021_Liu_Memory-efficient,2023_Yin_energy-absorption}. One of the essential elements in the design and manufacturing pipeline of lattice structures is triangulation, which discretizes the interconnected struts into triangles. Boundary representation, like these triangles, is important to the visualization, analysis, optimization, and fabrication steps in the pipeline~\cite{zou2023variational,li2023xvoxel,2024_meshing-for-FEM,zou2021length,su2020accurate,luo2023simple,wang2023computing,zou2013iso,zou2014iso,zou2021robust}.

Existing methods for lattice structure triangulation basically fall under two different strategies. The first strategy directly discretizes the struts into triangles~\cite{2017_factedDiamondMapping,2017_prefab-cell_HGM,2007_ChenYong_3DtextureMapping,2005_HongqingWang_STL-lattice, 2002_HongqingWang_conformal, 2006_chenYong_mesh-based, 2017_Chougrani_lightweight-triangulation}. In the second strategy, a lattice structure is first translated to other representation schemes~\cite{2020_Wu_CHoCC_triangulation,verma2020combinatorial,2021_Ding_STL-free,2022_Ma_MT-TPMS}, and then the triangulation is done on the translated version through, for example, the marching cube algorithm~\cite{2021_Stromberg_MC-TPMS}. 

The above two strategies can handle small-scale lattice structures but become insufficient when large-scale lattice structures with millions or billions of struts are considered. This is because lattice structures tend to have high geometric complexity, and a tremendous number of triangles need to be calculated and stored. Based on a back-of-the-envelope calculation, it would require $>$100TB of memory and take hours to weeks to instantiate a cube of 1m$^3$ volume with interconnected .1mm micro trusses~\cite{DARPA2020}. In addition, the diverse requirements of different applications give rise to the need for meshes with varying resolutions for the same lattice structure. This need is often met by starting over the triangulation algorithms again and again, which is unacceptable.

To address these limitations, this paper proposes a new representation scheme called meta-mesh for lattice structures. A meta-mesh is a collection of connected vertices, circular/elliptical arcs, and cylindrical/conical faces that make up a lattice's boundary (see Sec.~\ref{sec:method-overview}). It can be deemed as ``minimal" with regard to the number of vertices, arcs, and faces required to represent a lattice structure. More importantly, it can serve as a reusable base mesh for conveniently and efficiently triangulating lattice structures at arbitrary resolutions---only arc subdivision is needed, and there is no need for topology manipulation during triangulation. It can also generate a clean triangulation that avoids redundant triangles for a given error tolerance. Compared to the final fine triangular mesh, the number of elements in a meta-mesh can be significantly reduced. (Note that the meta-mesh is not a completely new notion, it has been occasionally and informally used in computer graphics~\cite{2019_schmidt_distortion}, and this paper extends it to the field of lattice structure modeling.) 

This paper also presents a CPU+GPU asynchronous pipeline to efficiently generate meta-meshes for billion-scale lattice structures. It only takes around 15 minutes to meta-mesh and then triangulate a billion-scale lattice structure (on Nvidia RTX 3090). Achieving such a high performance is by no means trivial, and the major challenges involved are as follows. 

\textbf{CPU-GPU data transfer bottleneck.} It is well known that data transfer is expensive and a major bottleneck in GPU computing~\cite{2018_Zhu_HBM-GPU}. For billion-scale lattice structures, this bottleneck is even more severe since their file size could be hundreds of GB or even TB~\cite{DARPA2020}. To mitigate this problem, this work designs a compressed representation of lattice structures to reduce the data to be transferred. Furthermore, an asynchronous meta-meshing pipeline is proposed to overlap data transfer and geometric computation, thereby hiding the data transfer latency.
    
\textbf{GPU-Lattice structural mismatch.} Lattice structures consist of highly interconnected struts, and therefore coupled geometric computing. By contrast, GPUs are designed to leverage parallelism, requiring computing tasks to be decoupled and independently processed on GPU cores. This mismatch between the data structure and the computing architecture presents challenges to parallel lattice structure triangulation. To solve this issue, we cast the coupled strut-strut intersections (which underlies triangulation and meta-meshing) as decoupled plane-strut intersections, which not only enable parallel computing but also are much easier to calculate.
    
\textbf{High warp divergence.} Almost all existing parallel computing algorithms in CAD~\cite{2008_Sara_NURBS,2012_Sara_Hausdorff,2014_Sara_3D} follow a thread-centric paradigm, which structures geometric computing around threads~\cite{2016_warpedslicer}. Quite often, this way of working causes unbalanced workloads among threads, warp divergence, and memory divergence, leading to low parallelism. By contrast, this work shifts from thread-centric to the very recent warp-centric paradigm, which structures geometric computing around warps~\cite{2019_Awad_BTree}. We achieve warp-centric meta-meshing by a novel GPU cache-aware data structure, a coalesced-access memory layout, and a workload-balanced scheduling method, which can distribute thread workloads evenly within the same warp, thereby achieving high parallelism.

\rev{}{
The main contributions of this paper are summarized as follows:
\begin{itemize}
    \item A meta-mesh representation scheme of lattice structures that is lightweight and reusable to attain multiresolution triangulation;
    \item A warp-centric GPU meta-meshing algorithm that can handle billion-scale lattice structures in minutes, with features of decoupled geometric calculation and low warp divergence; and
    \item A CPU+GPU asynchronous meta-meshing pipeline that can hide CPU/GPU data transfer latency and attain high triangulation throughput.
\end{itemize}
}

The remainder of this paper is organized as follows. Sec.~\ref{sec:related_work} reviews the literature, and Sec.~\ref{sec:preliminaries} gives a very brief introduction to GPU computing. Secs.~\ref{sec:methods} and~\ref{sec:triangulation} elaborate the meta-meshing method and the triangulation method, respectively. Validation of the method using a series of examples and comparisons can be found in Sec.~\ref{sec: results}, followed by conclusions in Sec.~\ref{sec:conclusion}.

\section{Related work}
\label{sec:related_work}
In this section, we briefly discuss literature related to lattice triangulation (Sec.~\ref{sec:lattice_triangulation}) and GPU parallel computation (Sec.~\ref{sec:GPU_parallel_computation}). Lattice triangulation methods can be categorized as direct and indirect (further divided into approximation- and isosurface extraction-based). The GPU parallel computation part summarizes relevant methods, including GPU memory access, warp-centric programming, and GPU-CPU cooperation.

\subsection{Lattice triangulation.}
\label{sec:lattice_triangulation}
\textbf{Direct triangulation.} This approach directly represents lattice structures as triangles. There are two kinds of methods in this category. The first involves replicating entirely meshed cell geometries (e.g., see~\cite{2017_factedDiamondMapping,2017_prefab-cell_HGM,2007_ChenYong_3DtextureMapping}). The second involves assembling meshed nodes and struts of cells separately (e.g., see~\cite{2005_HongqingWang_STL-lattice, 2002_HongqingWang_conformal, 2006_chenYong_mesh-based}), which helps avoid complex node-strut intersections. While current CAD tools can handle direct triangulation for small-scale lattice structures, processing billion-scale lattice structures using this approach presents challenges due to the model data explosion problem.

\textbf{Indirect triangulation through isosurface extraction.} The implicit modeling of lattice structures has been widely employed due to its compact representational advantages. With it comes research studies on the triangulation of implicit lattice structures. Several methods have been developed for this purpose, including the marching cube method~\cite{2021_Stromberg_MC-TPMS} and the marching tetrahedra method~\cite{2022_Ma_MT-TPMS}. However, it should be noted that the use of these algorithms may result in low-quality triangulations (e.g., redundant facets), giving rise to concerns regarding representational accuracy.

\textbf{Indirect triangulation via approximation.} A recently emerging lattice structure triangulation strategy is to mesh an approximating model of the lattice structure, instead of the original model. As involved in~\cite{2020_Wu_CHoCC_triangulation,verma2020combinatorial}, the convex hull nodes are used to approximate the spherical nodes of lattice structures, then a coarse mesh can be obtained from the approximated lattice structures. This can simplify the triangulation problem but results in error-uncontrollable and low-quality meshes.

In summary, direct triangulation methods suffer from the disadvantage of poor performance, while indirect triangulation methods (either based on iso-surface extraction or approximation) result in low-quality meshes. Additionally, the results from all methods are one-time and not reusable. To address these shortcomings, this paper introduces a meta-meshing approach (Sec.~\ref{sec:methods}).

\subsection{GPU parallel computation}
\label{sec:GPU_parallel_computation}
GPU computing has been extensively used in 3D modeling, including mesh simplification~\cite{2021_Mousa_Simplification}, point-set triangulation~\cite{2014_Cao_GPU-triangulation}, surface approximation~\cite{2022_Mousa_GPU-surface-approximation}, among others. Since the proposed method is highly related to GPU computing, we provide a summary of relevant GPU computing techniques and their recent advancements.

\textbf{GPU memory access.} 
Due to the high latency and low bandwidth of GPU global memory, memory access can cause performance bottlenecks as computations are stalled by memory transactions. Various techniques have been employed to address these bottlenecks, which include coalesced access, dedicated memories, and data compression. Coalesced access effectively mitigates memory transactions by modifying data layouts~\cite{2013_Cecilia_data-layout} or reorganizing threads~\cite{2014_Delbosc_reorginze-thread}. However, achieving coalescing is challenged when problems necessitate random access. Dedicated memories, such as constant memory~\cite{2018_Fang_Benchmark-GPUmemory} and texture memory~\cite{2010_Monakov_texture}, offer low access latency but are generally limited in capacity. Data compression is widely utilized to reduce both storage and access demands~\cite{2012_Boyer_Knapsack-problem}, but this approach introduces additional computational overhead.

\textbf{Warp-centric parallel programming.} 
“Warp-centric” denotes a programming paradigm that structures code around warps as units of computation~\cite{2023_Hijma_Optimization}. This approach is commonly utilized to reduce synchronization overhead~\cite{2018_Abdelfattah_warp-centric-sync} and minimize warp divergence~\cite{2010_Bos_warp-centric-divergence}. Implementing warp-centric programming involves reorganizing the code so that the work is assigned to warps instead of threads~\cite{2009_Bell_warp-centric-implement}. However, it’s essential to note that the warp-centric concept is inherently tied to the GPU hardware. Code optimized for a specific warp size or architecture may lack portability across diverse GPU platforms.

\textbf{GPU-CPU cooperation.} 
Research on cooperative GPU-CPU computing has been explored to improve overall application performance. The current focus remains on data transfer and work distribution between the GPU and CPU~\cite{2018_Raju_CPU-GPU-Cooperative}. Techniques for transferring encompass pipelining~\cite{2016_Yong_Histogram-CUDA} and data compression~\cite{2012_Boyer_Knapsack-problem}. Simultaneously, it is paramount to determine the most suitable device for each part of an application to run on~\cite{2018_Raju_CPU-GPU-Cooperative}. For the compute-intensive tasks of lattice triangulation, it's advisable to employ CPU-GPU cooperation to improve the performance.

In summary, the aforementioned techniques offer valuable inspiration to address the proposed challenges. However, it is important to note that there is presently no direct GPU algorithm accessible for lattice triangulation. Subsequent sections of this paper will integrate these techniques to solve the GPU meta-meshing problem.

\section{GPU Preliminaries}
\label{sec:preliminaries}
\begin{figure}[tb]
	\centering
	\includegraphics[width=0.45\textwidth]{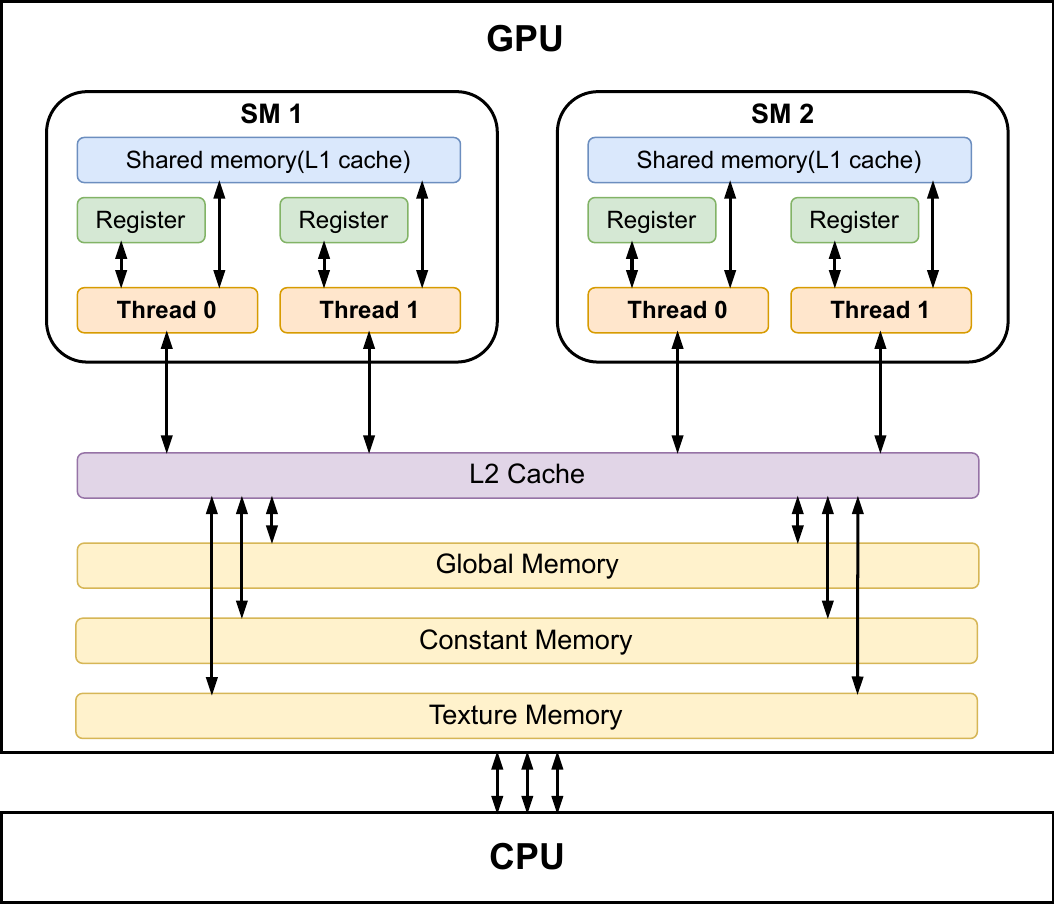}
	\caption{An overview of the GPU architecture.}
 \label{fig: gpu-structure}
\end{figure}

GPUs are typically structured with several stream multiprocessors (SMs), and each SM is structured with multiple cores, as shown in Fig.~\ref{fig: gpu-structure}. SMs have own dedicated local resources (e.g., L1 cache called shared memory, registers) and contain several blocks where each of them is a group of threads. All SMs have access to some public resources, such as global memory, and L2 cache positioned between L1 cache and global memory. Additionally, the GPU integrates constant memory and texture memory as read-only memory caches. 
\rev{}{Between the global memory and caches, data is transferred in unit called ``cache line", a quantity (typically 128 bytes) determined by GPU hardware.}

Practically, the GPU divides computational resources into warps in SMs, which contain multiple consecutive threads, e.g., 32 threads. Threads within a warp execute the same instruction simultaneously. Executing different instructions in a warp may result in warp divergence~\cite{2024_CUDA-programming}. Moreover, memory accesses are also issued per warp. 
\rev{If memory transactions cannot be coalesced into a single cache line, memory divergence occurs~\cite{2020_Wang_memory-divergence}.}{If memory transactions cannot be coalesced into a single cache line, the GPU has to perform multiple memory transactions to fetch relevant data from the global memory to the caches to support the calculations of all the threads in a warp.  During such memory transactions, irrelevant threads are forced to wait, resulting in low efficiency.  In the worst case, a memory transaction can only support a single thread’s data requirement, and all the other threads in the warp are waiting. This problem is referred to as memory divergence~\cite{2020_Wang_memory-divergence}.
}

To fully leverage the parallel capabilities of the GPU, an application should have the following features:
\begin{itemize}
    \item \textbf{Global memory access reduction.} Accessing global memory typically requires several hundred GPU cycles. Therefore, minimizing global memory access is critical to alleviate memory bottlenecks within GPU processing.
    \item \textbf{Memory and warp divergence reduction.} Both divergences would produce much idle time in GPU computing, leading to under-full utilization of cores and therefore decreasing GPU gains. For such problems, coalescing memory transactions and balancing the workload of warps are important for GPU performance.
\end{itemize}

\section{Meta-meshing}
\label{sec:methods}
\subsection{Meta-mesh definition and method overview}
\label{sec:method-overview}
\begin{figure}[t]
    \centering
    \subfigure[]{
        \includegraphics[width=0.4\textwidth]{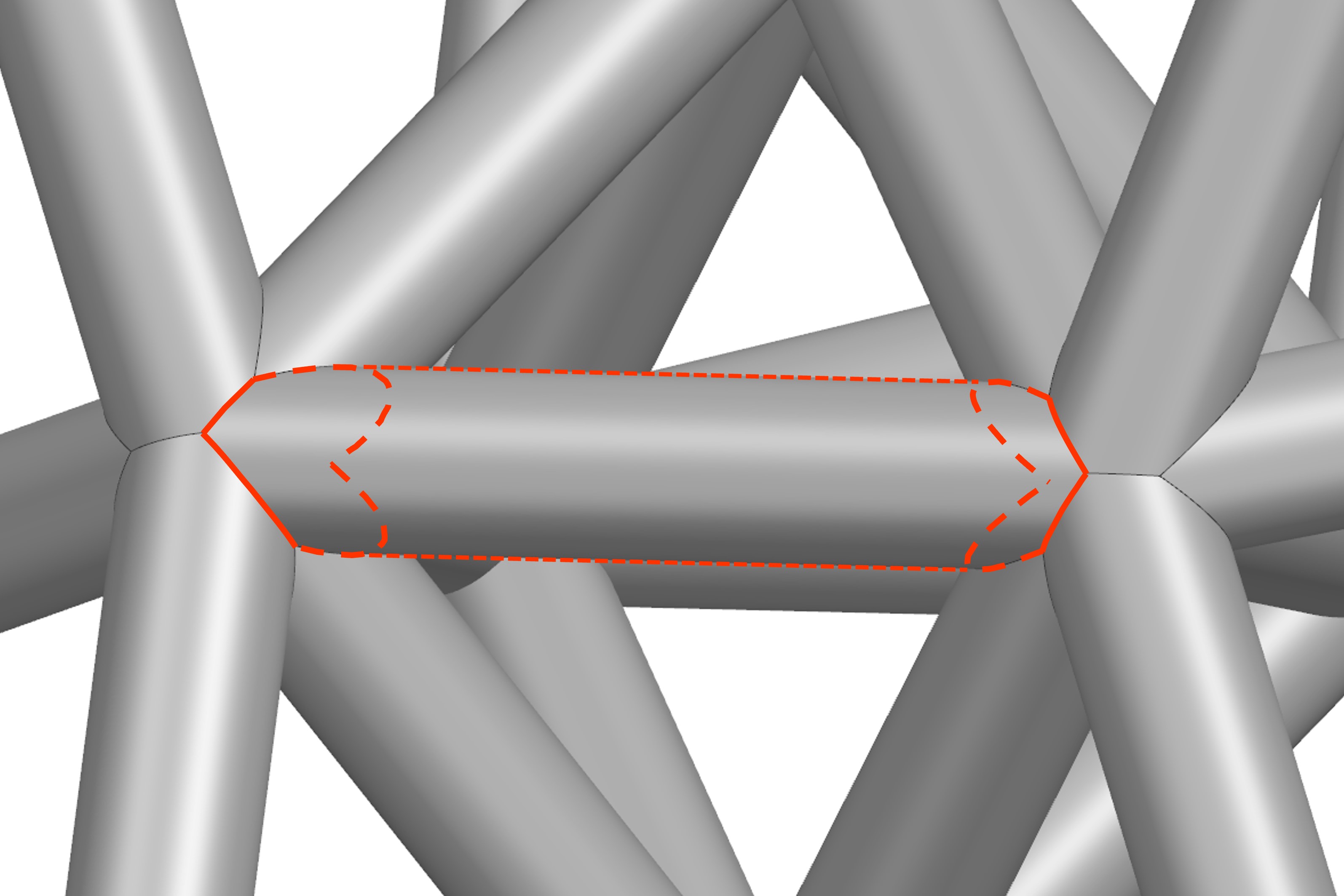}
    }
    \subfigure[]{
        \includegraphics[width=0.4\textwidth]{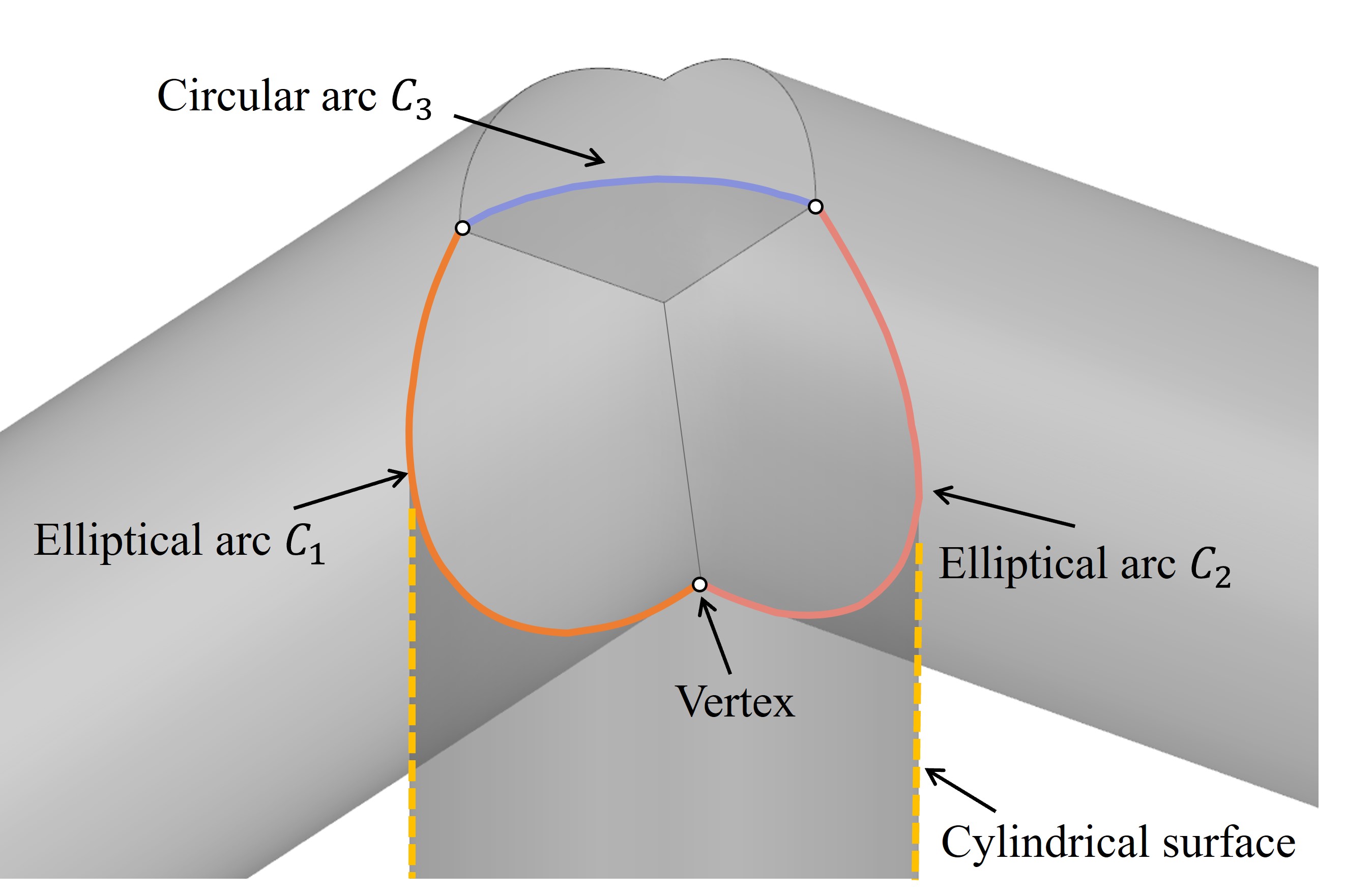}
    }
    \caption{Illustrations of meta-meshes: (a) a single face; and (b) an arc loop.}
    \label{fig: meta-mesh}
\end{figure}

A meta-mesh is the collection of vertices, circular/elliptical arcs, and trimmed cylindrical/conical surfaces, which make up the boundary of lattice structures (Fig.~\ref{fig: meta-mesh}a). Each strut intersects with several adjacent struts, resulting in trimmed boundary surfaces. These intersection edges on the same side of a strut, which are circular or elliptical arcs, are sequentially interconnected, forming an arc loop as shown in Fig.~\ref{fig: meta-mesh}b. The vertices are located at the junctions of arcs. In such a case, lattice structures can be represented with a minimal number of geometric entities like vertices, arcs, and surfaces. With explicitly defined boundaries, the triangulation of lattice structures at an arbitrary resolution can be efficiently obtained through arc subdivision.

The notion of meta-mesh is simple by itself since it only decomposes a lattice structure into bounded surfaces according to the natural strut-strut intersections. However, meta-meshing a lattice structure is not trivial. This is because each arc on a strut is derived from the strut-strut intersection and needs to solve the overlap with other arcs to determine the connected vertices. For a billion-scale lattice structure, it will take days to weeks to generate the meta-mesh on an ordinary CPU, which is unacceptable. 
Accordingly, utilizing GPU to process large-scale meta-mesh is preferred. 
As already noted in the Introduction section, there are three major GPU meta-meshing challenges involved in GPU meta-meshing, and new developments are needed. These developments are summarized below:
\begin{itemize}
    \item \textbf{High-throughput lattice structure data transfer.} 
    Two approaches are utilized to improve the transfer speed. First, the lattice structures are compressed to solve the CPU-GPU transfer bottleneck. Second, an asynchronous pipeline is designed to hide the latency between GPU, CPU and disk. These two will be detailed in Sec.~\ref{sec:transferring_approach}.
    \item \textbf{Decoupled lattice structure computation.}
    To overcome GPU-Lattice structural mismatch, lattice structure decoupling is proposed to make GPU meta-meshing available. By introducing auxiliary planes, interconnected struts are decoupled into distinct elements so that meta-meshing can be parallelized on the GPU. See Sec.~\ref{sec: decoupling} for more details.
    \item \textbf{Warp-centric lattice structure meta-meshing.} 
    To implement warp-centric meta-meshing of lattice structures, a coalesced-access memory layout is designed to optimize memory transactions. Then, the warp-centric task scheduling is used to maximize GPU efficiency and performance gains. These related methods will be illustrated in Secs.~\ref{sec: coalesced-memory-access} and~\ref{sec: warp-scheduling}.
\end{itemize}

This work focuses on lattice structures with cylindrical and conical struts; these struts also have tangential relationships with nodal spheres. Lattice structures of this kind are among the most commonly used ones~\cite{2017_Chougrani_lightweight-triangulation}.

\subsection{Data transfer of billion-scale lattice structures}
\label{sec:transferring_approach}
Meta-meshing billion-scale lattice structures inevitably suffer from data transfer bottlenecks. This leads to the GPU spending more than half of the time waiting for data transfers rather than computing.
Therefore, a compressed lattice structure representation with controlled errors is proposed to accelerate data transfers. Besides, a heterogeneous asynchronous pipeline is designed to maximize bus bandwidth and throughput.

\subsubsection{Compressive lattice structure representation}
\label{sec: compression}
\begin{figure}[t]
	\centering
	\includegraphics[width=0.4\textwidth]{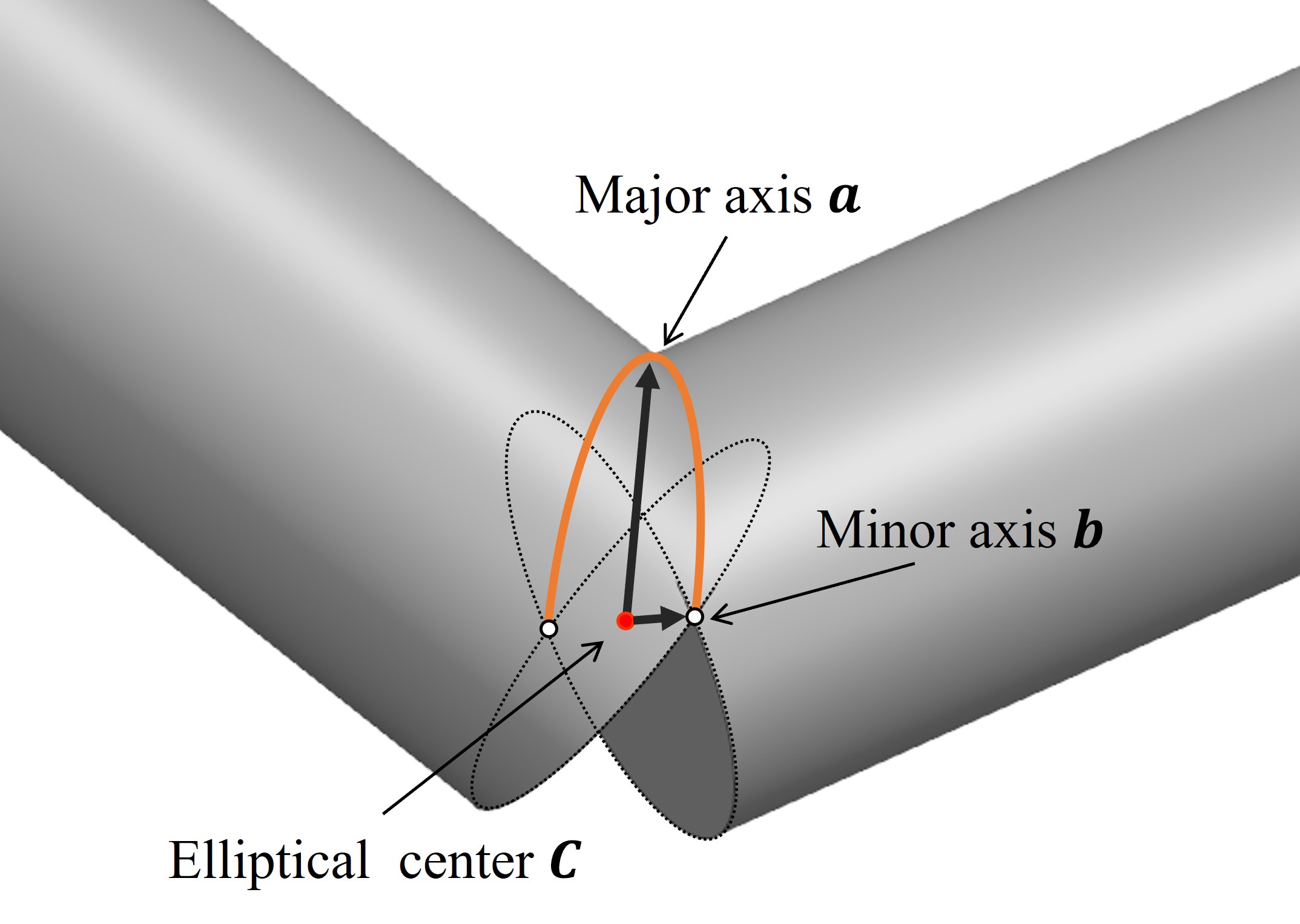}
	\caption{Illustration of an elliptical arc generated by intersecting two struts.}
 \label{fig: ellipse}
\end{figure}

When two struts intersect at a node, the intersection curve is a truncated elliptical arc as shown in Fig.~\ref{fig: ellipse}, which can be expressed as:
\begin{equation}
\begin{split}
  \begin{cases}
    x &= \boldsymbol{a}_xsin(t) + \boldsymbol{b}_xcos(t) + \boldsymbol{c}_x\\
    y &= \boldsymbol{a}_ysin(t) + \boldsymbol{b}_ycos(t) + \boldsymbol{c}_y\quad\forall t\in [t_1, t_2]\\
    z &= \boldsymbol{a}_zsin(t) + \boldsymbol{b}_zcos(t) + \boldsymbol{c}_z
  \end{cases}
\end{split}
\label{elliptical_curve_equ}
\end{equation}
where the vectors $\boldsymbol{a, b}$ are the ellipse's \rev{major}{semi-major} axis and \rev{minor}{semi-minor} axis, respectively. \rev{C is the center of the ellipse located at the node.}{$\boldsymbol{c}$ is the center of the ellipse.} $t_1$ and $t_2$ are the parameter values of the arc's two ends.

To completely store this arc's information, 11 parameters are needed (9 for three vectors, 2 for the parameter domain). Assuming each parameter occupies a 4-byte floating-point number in GPU, the entire curve demands 44 bytes. For billion-scale lattice structures, there are billions of intersection curves, and the allocation for such large-scale geometric data in memory-limited GPU is unfeasible. 

To address this issue, a compressed storage scheme is proposed, which is inspired by the Fixed-Point Representation method~\cite{2006_Lindstrom_Integer-compression}. This compression technique quantizes float-point values within a range onto a uniform integer grid for efficient predictive compression. Based on this compression scheme, the vectors $\boldsymbol{a, b}$ \rev{}{and $\boldsymbol{c}$} in Eq.~\eqref{elliptical_curve_equ} are first converted to spherical coordinates, which is described by three parameters of radial distance $r$, the polar angle $\theta$, and the azimuthal angle $\varphi$. The two angle parameters inherently have fixed ranges ($[0,\pi]$ for polar angle, $[0,2\pi]$ for azimuthal angle). For radial distance $r$, it can be easily bounded by its minimum and maximum values~(\rev{$R \leq r \leq 4R$, $R$ is the cylinder radius}{the length of the semi-major axis lies in the interval $[R_{min}, 4R_{max}]$, the semi-minor axis in $[0.1R_{min}, R_{max}]$ and the center in $[0, R_{max}]$, where $R_{min}$ and $R_{max}$ are minimum and maximum node radii, respectively})\footnote{\rev{}{It is possible to have the lengths of the ellipse's semi-major axes exceed the minimum and maximum values used in this work.  If such cases occur, we turn to the non-compressed scheme (which of course would affect the overall computational efficiency).  Fortunately, such cases are rare, and they account for only a small percentage out of all cases, less than 0.01\% based on our statistics.}}. As a result, all these parameters can be coded into one or more integers.

\begin{figure}[htbp]
\centering
\subfigure[]{
    \includegraphics[width = 0.22\textwidth]{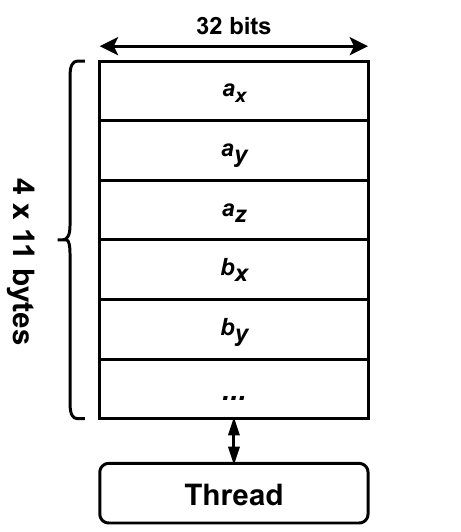}
}
\subfigure[]{
    \includegraphics[width = 0.22\textwidth]{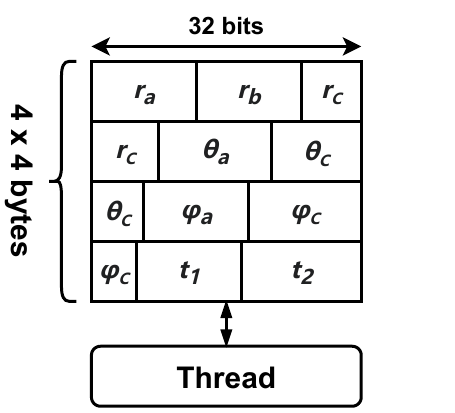}
    \label{fig: compress_b}
}
\caption{The storage of arcs in GPU global memory: (a) uncompressed storage; and (b) compressed storage.}
\label{fig: compressed-structure}
\end{figure}

\rev{Furthermore, some parameters do not need to be stored, such as the elliptical center $\bm{O}$ and the radial distance of the minor axis $r_b$. In the case of the elliptical center $\bm{O}$, it is located at the node where the struts intersect, which has already been stored in the memory. For the radial distance of the minor axis $r_b$, its length is the radius $R$. After removing redundant parameters, the compression scheme is illustrated in Fig.~\ref{fig: compressed-structure}b.}
{Furthermore, some parameters do not need to be stored, such as the polar angle $\theta_b$ and the azimuthal angle $\varphi_b$ of the semi-minor axis $\boldsymbol{b}$. Since the semi-minor axis is perpendicular to both the strut's direction and the semi-major axis, $\theta_b$ and $\varphi_b$ can be simply obtained by the cross-product. After removing redundant parameters, the compression scheme is illustrated in Fig.~\ref{fig: compressed-structure}b.}

\textbf{Error analysis and control.} 
Since the proposed compression method is lossy, the number of bits, used to code a floating-point number, needs to be optimally determined to control compression error.
For any point $\boldsymbol{p}$ on the arc with parameter $t$, there is a corresponding point $\boldsymbol{p}'$ after compression. The error \rev{}{$E$} between these two points can be measured by the distance between $\boldsymbol{p}$ and $\boldsymbol{p}'$:
\begin{equation}
\begin{split}
    E =& ||\boldsymbol{p}(t) - \boldsymbol{p}'(t)||\\
      =& ||(\boldsymbol{a}-\boldsymbol{a}')sin(t) + (\boldsymbol{b} - \boldsymbol{b}')cos(t) + (\boldsymbol{c} - \boldsymbol{c}')||
\end{split}
\end{equation}
\rev{where $\boldsymbol{a}$ and $\boldsymbol{b}$ represent the major and minor axes of the uncompressed arc. Correspondingly, $\boldsymbol{a}'$ and $\boldsymbol{b}'$ are the major and minor axes after compression. To get the error of $(\boldsymbol{a} - \boldsymbol{a}')$ and $(\boldsymbol{b} - \boldsymbol{b}')$, we estimate the error $D$ induced by compressing a single vector as follows: }
{where $\boldsymbol{a}$, $\boldsymbol{b}$ and $\boldsymbol{c}$ represent the elliptical parameters of the uncompressed arc. Correspondingly, $\boldsymbol{a}'$, $\boldsymbol{b}'$ and $\boldsymbol{c}'$ are the parameters after compression. To get the error of $(\boldsymbol{a} - \boldsymbol{a}')$, $(\boldsymbol{b} - \boldsymbol{b}')$ and $(\boldsymbol{c} - \boldsymbol{c}')$, we estimate the error induced by compressing a single vector as follows: }
\rev{}{}
\begin{equation}
\begin{split}
 & ||\boldsymbol{a} - \boldsymbol{a}'||\\
 =& \sqrt{(a_x - a_x')^2 + (a_y - a_y')^2 + (a_z - a_z')^2}\\
 =& \sqrt{r^2+r'^2-2rr'(sin\theta sin\theta'cos(\varphi-\varphi') + cos\theta cos\theta')}\\
 <& \sqrt{(r-r')^2 + 2rr' - 2rr'cos(\theta-\theta')cos(\varphi-\varphi')}\\
\end{split}
\end{equation}
where $(a_x,a_y,a_z)$ and $(a_x',a_y',a_z')$ are Euclidean coordinates of $\boldsymbol{a}$ and $\boldsymbol{a}'$. Then they are substituted by spherical coordinates $(r, \theta, \varphi)$ and $(r',\theta', \varphi')$. 
Assuming $n$ bits for coding radial distance $r$, $m$ bits for coding polar angle $\theta$, and $m+1$ bits for the azimuthal angle $\varphi$. 
\rev{Because $r$ typically ranges in $[R,4R]$, when encoding $r$ using $n$ bits, $(r-r')$ has an error of $(4R-R)/2^{n+1}$.}{Because the semi-major axis' length is restricted to $[R_{min},4R_{max}]$, when encoding $r$ and $r'$ using $n$ bits, $(r-r')$ has an error of $(4R_{max}-R_{min})/2^{n+1}$.} For $\theta\in[0,\pi]$ and $\varphi\in[0,2\pi]$, they are respectively encoded using $m$ and $m+1$ bits. So $(\theta-\theta')$ and $(\varphi-\varphi')$ both have the error of $\pi/2^{m+1}$. \rev{As a result, the maximum value of $D$ is given by:}{As a result, the maximum value of $||\boldsymbol{a}-\boldsymbol{a}'||$ is estimated as:}
\begin{equation}
\begin{split}
||\boldsymbol{a}-\boldsymbol{a}'||_{max} < R_{max}\sqrt{\cfrac{16}{2^{2n+2}} + \cfrac{32\pi^2}{2^{2m+2}}}
\end{split}
\end{equation}
\rev{}{The error of $(\boldsymbol{b}-\boldsymbol{b}')$ and $(\boldsymbol{c}-\boldsymbol{c}')$ can be estimated in the same way.}

Using the distance between vectors before and after compression, the max error of the elliptical arc can be estimated as:
\begin{equation}
\label{Equ: error-result}
\begin{split}
    E \leq& ||\boldsymbol{a}-\boldsymbol{a}'||_{max}sin(t) + ||\boldsymbol{b}-\boldsymbol{b}'||_{max}cos(t) + ||\boldsymbol{c}-\boldsymbol{c}'||_{max}\\
    \leq& R_{Max}\left(\sqrt{\cfrac{17}{2^{2n+2}} + \cfrac{32\pi^2}{2^{2m+2}}} + \sqrt{\cfrac{1}{2^{2n+2}} + \cfrac{32\pi^2}{2^{2m+2}}}\right)
\end{split}
\end{equation}

For triangulation applications, a max error less than 0.001\rev{$R$}{$R_{max}$} is sufficient according to our experience. By substituting 0.001\rev{$R$}{$R_{max}$} to Eq.~\eqref{Equ: error-result}, we can get 12 bits for the radial line $r$, \rev{14}{15} bits for the polar angle $\theta$ and \rev{15}{16} bits for the azimuthal angle $\varphi$ to compress a vector. For the other information like parameter range $t_1$ and $t_2$, they are correspondingly compressed in \rev{13}{15} bits. After removing redundant parameters like $\theta_b$ and $\varphi_b$, the storage of an individual arc can be compressed from 352 bits to \rev{96}{128} bits.

\subsubsection{Heterogeneous asynchronous pipeline}
When the meta-mesh of a lattice structure is generated in the GPU, it is transferred to the CPU for the subsequent triangulation task. There are good reasons for using this CPU+GPU heterogeneous asynchronous pipeline. First, the triangular mesh of lattice structures has a large amount of geometric data; if using GPU for the triangulation, transferring so many triangles from GPU to CPU cannot be done in a short time. Second, the use of meta-meshes can make the triangulation simple and fast, thereby making it feasible to do triangulation on the CPU. Therefore, using GPU only for meta-meshing and CPU for triangulation improves the entire performance and avoids the bottleneck of CPU-GPU data transfers.

\begin{figure}[htbp]
\label{fig: pipeline}
\centering
\subfigure[]{
    \includegraphics[width=0.4\textwidth]{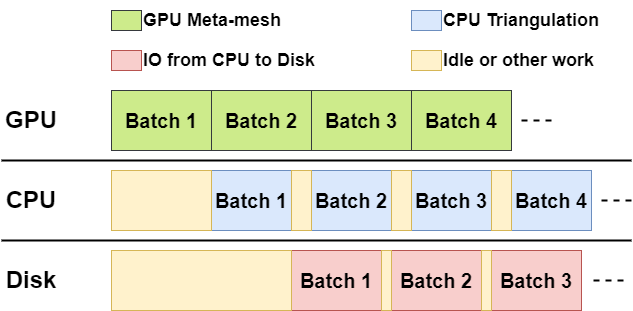}
}
\subfigure[]{
    \includegraphics[width=0.46\textwidth]{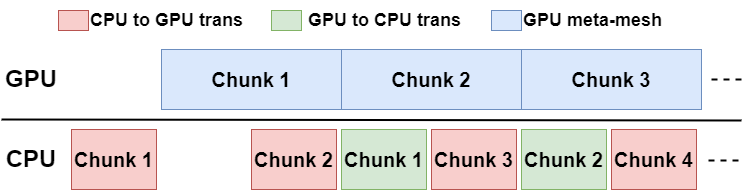}
}
\caption{The asynchronous pipeline: (a) GPU-CPU-Disk asynchronous execution; and (b) GPU-CPU asynchronous transfers.}
\end{figure}

To collaboratively utilize CPU, GPU, and IO devices, a heterogeneous asynchronous pipeline is designed, which contains two levels of asynchrony. The first asynchronously executes the meta-meshing of lattice structures in GPU, the triangulation based on meta-mesh in CPU, and the final file writing. With the batch-by-batch processing manner, when the CPU triangulates the current batch, the GPU can meta-mesh the next batch while the I/O bus can write the last batch to the disk, as shown in Fig.~\ref{fig: pipeline}a. Another level of asynchrony exists in data transfers between GPU and CPU. A larger batch is further subdivided into small chunks suitable for GPU computation. While the GPU is processing the current chunk, the next chunk is transferred to the GPU, and the processing result of the last chunk is moved out GPU. This two-level way of working can effectively hide GPU-CPU data transfer latency, as shown in Fig.~\ref{fig: pipeline}b.

To effectively run the above pipeline, some parameters need to be set properly. The batch needs to have a large size so that it can be written to disk in blocks, but it is also limited by the I/O buffer size. To solve this challenge, we provide the following equation to set a proper batch size:
\begin{equation}
\label{Equ: chunk-size}
\begin{split}
    N_{batch} = \cfrac{S_{buffer}}{S_{triangle}*N_{strut}}
\end{split}
\end{equation}
where $N_{batch}$ is the number of struts in a batch, $S_{buffer}$ is the buffer size, $S_{triangle}$ is the triangle size, and $N_{strut}$ is the number of triangles on a strut. Considering an example, an average of 20 triangles per strut ($N_{strut}=20$), 50 bytes per triangle ($S_{triangle}=50$), and a 4MB buffer size ($S_{buffer} = 4,000,000$). Based on Eq.~\eqref{Equ: chunk-size}, a batch can contain 4000 struts. 
\rev{To hide the latency of CPU-GPU transfers, dividing each batch into 4 chunks is sufficient, which contains 1000 struts per chunk.}{To hide the latency of CPU-GPU data transfer, we further divide each batch into 4 chunks, each of which contains 1000 struts. Like most GPU computation problems, there is no principled way to determine the chunk size, and we set it to 4 through empirical studies.  We conducted experiments on a lattice structure dataset and examined different chunk sizes (from 1 to 10), and eventually found that dividing each batch into 4 chunks gave the best performance.}

\subsection{Warp-centric GPU meta-meshing}
\label{sec:warp-centric_approach}
To solve the GPU-Lattice structural mismatch challenge, a lattice structure decoupling scheme is proposed, which converts interconnected lattice structures into decoupled GPU-suitable parts. Then, coalesced memory layouts and optimized warp-centric scheduling are presented to reduce memory and warp divergence.

\subsubsection{Decoupled lattice structure computation}
\label{sec: decoupling}
The computation of meta-mesh involves determining the intersections between a strut and its adjacent struts, which generate elliptical arcs. Since arcs can overlap, their computation is interrelated, which is not in line with the requirement of independent tasks for GPU threads. In this work, auxiliary planes are introduced to decouple the computation on adjacent struts.
By replacing adjacent struts with planes, elliptical arcs can be computed in parallel through strut-plane intersections.

\begin{figure}[tb]
    \centering
    \subfigure[]{
        \includegraphics[width = 0.4\textwidth]{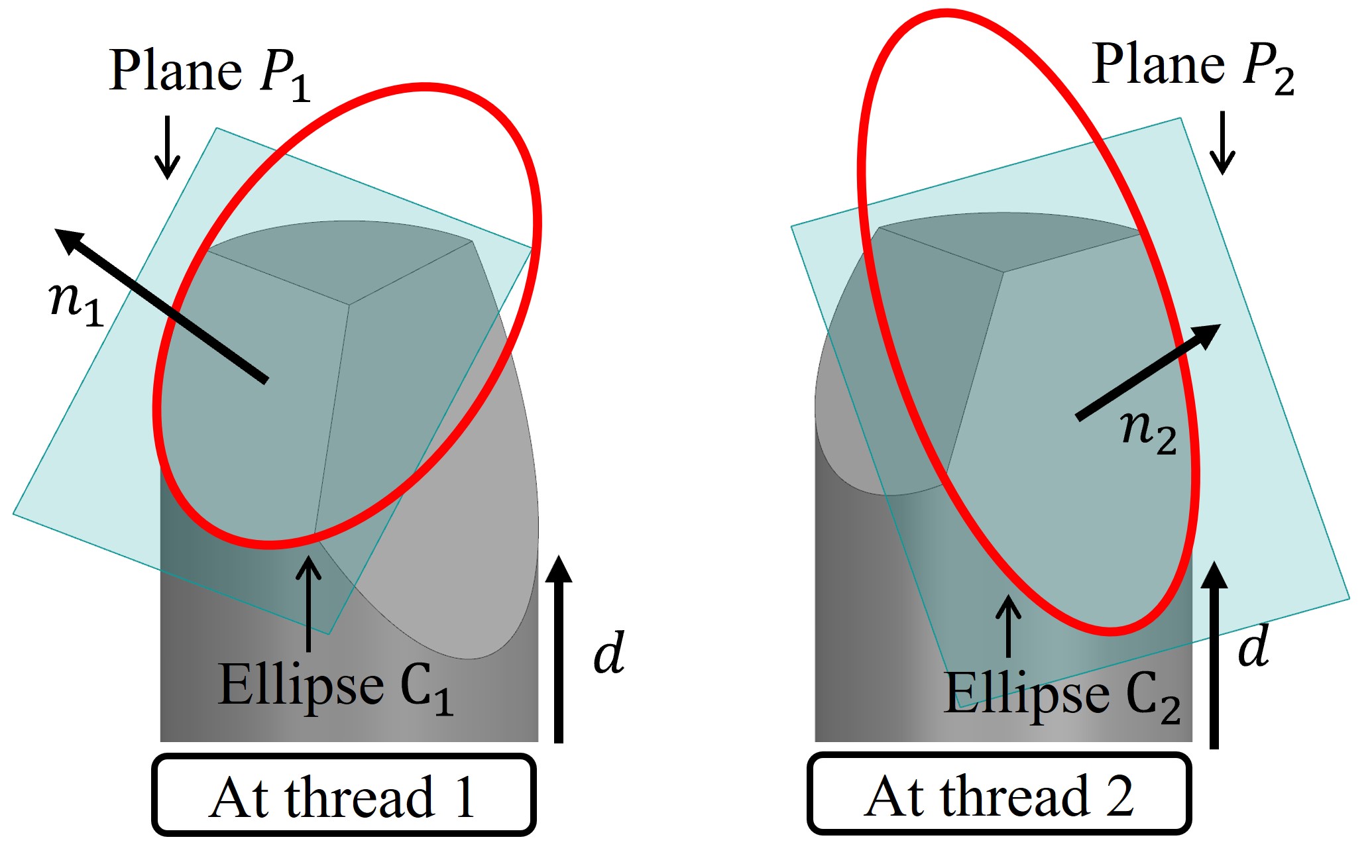}
    }
    \subfigure[]{
        \includegraphics[width = 0.4\textwidth]{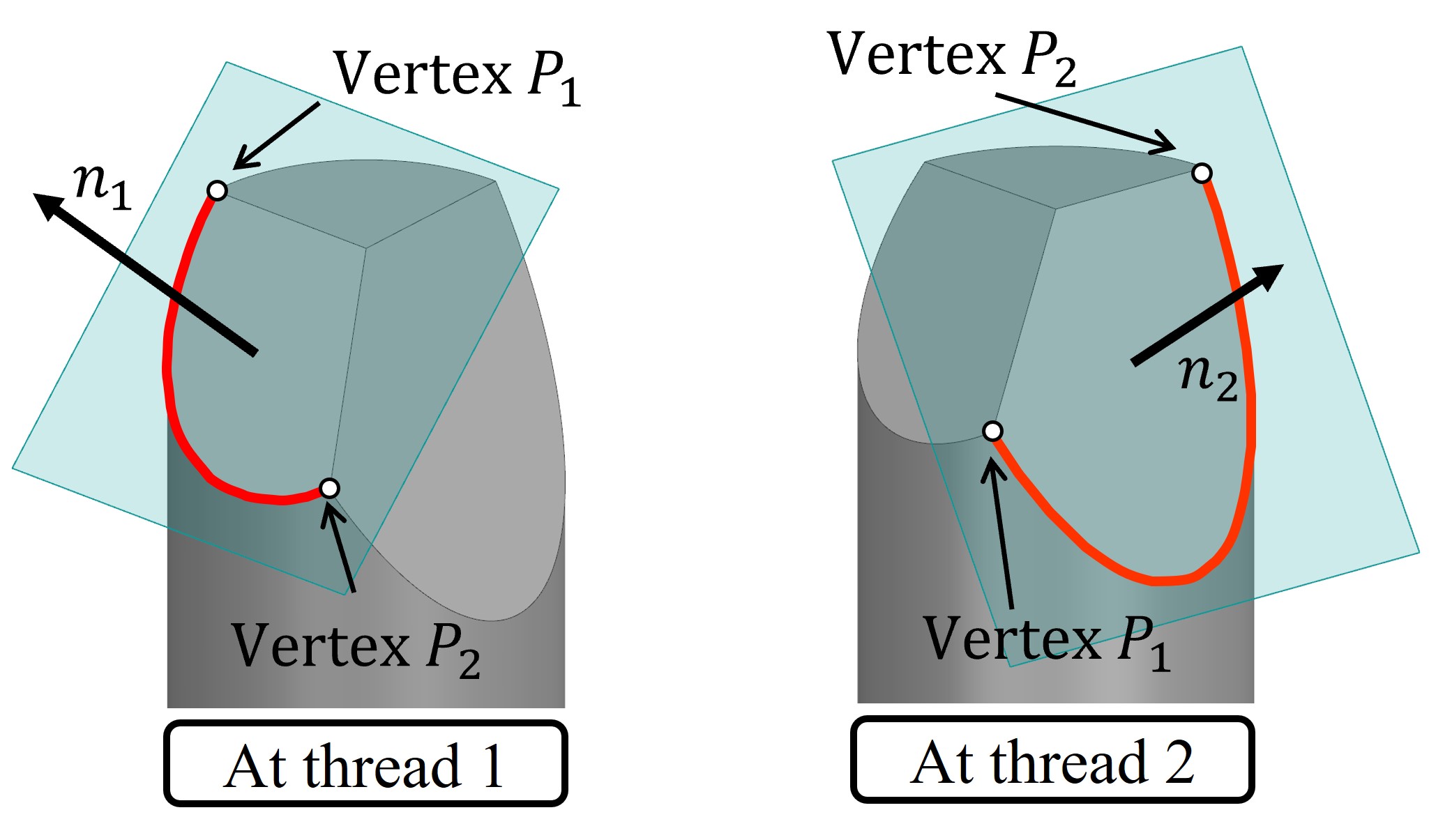}
    }
    \caption{The decoupled computing scheme for lattice structures: (a) strut-plane intersection for ellipse computation; and (b) ellipse-plane intersection to get arc loop.}
    \label{fig: strut-notations}
\end{figure}

\rev{Suppose a strut is defined as connected vertices $(V_0$, $V_1)$ and its direction is defined as a unit vector $\boldsymbol{d}$. The strut $(V_0, V_1)$ intersects with other struts at $V_0$, generating several intersected planes as $P_0,P_1,...,P_k$.  Correspondingly, the normal vectors of planes are $\boldsymbol{n}_0, \boldsymbol{n}_1,..., \boldsymbol{n}_k$. For a plane $P_i$, its intersected ellipse $C_i$ is given by:}{Suppose a strut $S_i$ is defined by its two end vertices $\bm{v}_i^{0}, \bm{v}_i^{1}$ and the corresponding radii $r_i^{0}, r_i^{1}$. Further assume that this strut intersects with its adjacent struts at $\bm{v}_i^{0}$. Following from Rossignac's work~\cite{2018_Gupta_Quador}, the intersection curve of two struts is a planar conic section, and the plane carrying the intersection curve can be easily calculated from the struts' vertices and radii. In this work, we precompute such planes using the GPU and the lattice’s topology graph (i.e., an adjacency list in the GPU). Let the plane for two connecting struts $S_i, S_j$ be denoted by $P_{i,j}$ (to be called the auxiliary plane). We can then decouple the strut-strut intersection problem between $S_i, S_j$ into two simpler and independent strut-plane problems, i.e., between $S_i, P_{i,j}$ and between $S_j, P_{i,j}$. Finding the intersection (i.e., an ellipse) between a plane and a cylindrical/conical strut is an elementary geometry problem, and the equations for calculating the center $\bm{o}_{i,j}$, the semi-major axis $\bm{a}_{i,j}$, and the semi-minor axis $\bm{b}_{i,j}$ are as follows~(Chapter 1 of~\cite{brannan2011geometry}):}
\begin{equation}
    \begin{split}
        &\boldsymbol{o}_{i,j} = \cfrac{1}{2}(\boldsymbol{E}_1 + \boldsymbol{E}_2) \\
        &\boldsymbol{a}_{i,j} = \cfrac{1}{2}(\bm{E}_1 - \bm{E}_2) \\
        &\boldsymbol{b}_{i,j} = \lambda (\boldsymbol{a}_{i,j}\wedge\boldsymbol{n}_{i,j}) \\
with\quad  &\bm{E}_1 = \cfrac{\bm{n}_{i,j} \cdot (\bm{p}_{i,j}-\bm{F}_1)}{\bm{n}_{i,j} \cdot Rot(\bm{d_i^{\prime}},+\alpha)\bm{d}_i}\times Rot(\bm{d_i^{\prime}}, +\alpha)\bm{d}_i + \bm{F}_1\\
            &\bm{E}_2 = \cfrac{\bm{n}_{i,j} \cdot (\bm{p}_{i,j} - \bm{F}_2)}{\bm{n}_{i,j} \cdot Rot(\bm{d_i^{\prime}}, -\alpha)\bm{d}_i}\times Rot(\bm{d_i^{\prime}}, -\alpha)\bm{d}_i + \bm{F}_2 \\
            &\bm{F}_1 = \bm{v}_i^0 + Rot(\bm{d_i^{\prime}},\alpha - \cfrac{\pi}{2})\bm{d}_i \times r_i^0\\
            &\bm{F}_2 = \bm{v}_i^0 + Rot(\bm{d_i^{\prime}},\cfrac{\pi}{2} - \alpha)\bm{d}_i \times r_i^0\\
            &\lambda  = \sqrt{1 - (\bm{a}_{i,j} \cdot \bm{d}_i)^2 / (cos\alpha \times \|\bm{a}_{i,j}\|)^2}\\
    \end{split}
\end{equation}
\rev{where ``$\wedge$" is the symbol used for cross product. $\boldsymbol{a}_i$ and $\boldsymbol{b}_i$ correspond to the major and minor axes of the ellipse, with the center $O_i$. $R$ is the strut radius at $V_0$, which is an endpoint of the strut.}{where operators ``$\cdot$" and ``$\wedge$" denote dot and cross product, respectively. $\bm{n}_{i,j}$ and $\bm{c}_{i,j}$ are the orientation and position of the plane $p_{i,j}$, respectively. $Rot(\bm{d^{\prime}}, \alpha)$ is the matrix rotating a vector around axis $\bm{d}^{\prime}$ by angle $\alpha$. $\alpha$ is strut's conical angle given by $arccos\left((r_i^1 - r_i^0)/||\bm{v}_i^1 - \bm{v}_i^0||\right)$~(For cylindrical struts, $\alpha$ is 0). $\bm{d}_i$ is the normalized direction vector from $\bm{v}_i^1$ to $\bm{v}_i^0$, and $\bm{d_i^{\prime}}$ is the normalized version of $\bm{d}_i \wedge \bm{n}_{i,j}$.} 
By looping through intersected planes, all ellipses can be obtained at the end section of the strut. This process can be effectively parallelized on the GPU by assigning an ellipse into a thread (See Fig.~\ref{fig: strut-notations}a).

After determining the intersected ellipses, there are overlaps among ellipses on the strut. So one must compute the elliptical arc that is not overlapped by other ellipses. That is to say, the connected vertices between ellipses need to be calculated. To avoid time-consuming ellipse-ellipse intersections, an ellipse is replaced by the auxiliary plane, and the connected vertices can be obtained by the ellipse-plane intersection. Assuming ellipse \rev{$C_i$}{$C_{i,j}$} overlaps with ellipse \rev{$C_j$}{$C_{i,k}$}, we can get the arc range $t$ on ellipse \rev{$C_i$}{$C_{i,j}$} with following equations:
\begin{equation}
    \label{equ: arc-range}
    \begin{split}
    \begin{cases}
        \boldsymbol{v} = \boldsymbol{a}_{i,j}sin(t) + \boldsymbol{b}_{i,j}cos(t) + \boldsymbol{o}_{i,j}\\
        \boldsymbol{n}_{i,k} \cdot (\boldsymbol{v} - \boldsymbol{o}_{i,k}) \leq 0
    \end{cases}
    \end{split}
\end{equation}
\begin{equation}
\label{equ: overlap-vertices}
\begin{split}
 cos^{-1}\cfrac{C}{\sqrt{A^2 + B^2}} \leq t &+ tan^{-1}\cfrac{B}{A} \leq 2\pi - cos^{-1}\cfrac{C}{\sqrt{A^2 + B^2}} \\
 where \quad &A = \boldsymbol{n}_{i,k} \cdot \boldsymbol{a}_{i,j} \\
             &B = \boldsymbol{n}_{i,k} \cdot \boldsymbol{b}_{i,j} \\
             &C = \boldsymbol{n}_{i,k} \cdot (\boldsymbol{o}_{i,k} - \boldsymbol{o}_{i,j})\\
\end{split}
\end{equation}
where \rev{$\bm{v}_i$}{$\bm{v}$} is an arbitrary point on ellipse \rev{$C_{i}$}{$C_{i,j}$}, which does not overlap with \rev{$C_j$}{$C_{i,k}$}. \rev{$\bm{n}_j$}{$\bm{n}_{i,k}$} is the normal of the plane where \rev{$C_j$}{$C_{i,k}$} lies on. With inter-thread communication (e.g., ``shuffles"), \rev{$\boldsymbol{n}_j$}{auxiliary planes} can be broadcasted to other threads for the ellipse-plane intersection, and the intersected vertices can be computed for each duet of ellipses.

From Eq.~\eqref{equ: overlap-vertices}, the elliptical arc can be obtained, which connected other arcs at vertices $P_1$ and $P_2$, as shown in Fig.~\ref{fig: strut-notations}b. By merging these arcs into an arc loop, the meta-mesh is acquired and then will be written to global memory in a coalesced way, as will be explained in Sec.~\ref{sec: coalesced-memory-access}.

\subsubsection{Coalesced-access memory layout}
\label{sec: coalesced-memory-access}
The meta-meshing results, a mass of arc loops, have to be temporarily stored in global memory. Usually, the arcs within the same warp are positioned at different addresses. Accessing these arcs leads to multiple memory transactions per warp, which causes serious memory latency. To solve this problem, a coalesced memory layout is proposed to coalesce the memory transactions of warps for storing arcs.

\begin{figure}[t]
	\centering
	\includegraphics[width=0.40\textwidth]{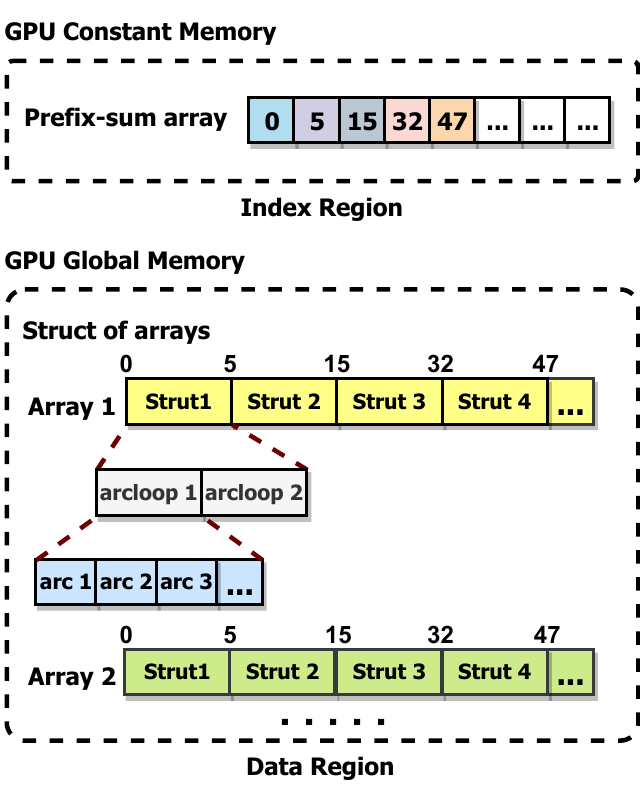}
	\caption{The memory layout design for storing meta-meshes.}\label{fig: memory-layout}
\end{figure}

The coalesced memory layout is handled by an index region and a data region. The index region is a prefix-sum array, which records the starting address of the meta-mesh for each strut. For example, the index region in Fig.~\ref{fig: memory-layout} is [0, 5, 15, 32, 47...]. It means the first arc of strut 1 is at position 0 on the data region and strut 2 is at position 5 and so on. By subtracting the index from its next index, the number of arcs of a strut can be easily obtained. 

The data region stores the meta-mesh consecutively, to make the warp access to arcs can be as coalesced as possible. The two arc loops of each strut are stored contiguously in memory. An arc loop consists of several arcs, which are sequentially connected. The arc is the smallest unit in the meta-mesh and is organized as Structure of Arrays (SoA) in the data region. In this organization, the data region holds several one-dimensional arrays of the same length, and then each arc is split into these arrays. To make memory access coalescing in a cache line, the size of each arc element is defined by:
\begin{equation}
\label{equ: element}
\begin{split}
    S_{element} &= \cfrac{S_{cacheline}}{warpsize} \\
\end{split}
\end{equation}
where $S_{element}$ is the size of each element in the SoA, and $S_{cacheline}$ is the cache line size in the GPU. 
For example, given a cache line size of 128 bytes and a warp size of 32, it is straightforward to determine that each element is 4 bytes in size. Consequently, when threads in a warp access a 4-byte arc element in the data region, it allows for the coalescing of adjacent 32 elements into a single cache line.

In memory, the index region is small (a chunk of 1024 struts at most uses 8KB) but is frequently accessed by the warp. Therefore, the prefix-sum array in the index region can be saved in low-latency constant memory, which can reduce memory latency in warps. Also, the prefix-sum array can be efficiently obtained in GPU through parallel scanning algorithms~\cite{2007_Harris_Parallel-scan}.

\subsubsection{Optimized warp scheduling}
\label{sec: warp-scheduling}
\begin{figure}[tb]
	\centering
	\includegraphics[width=0.48\textwidth]{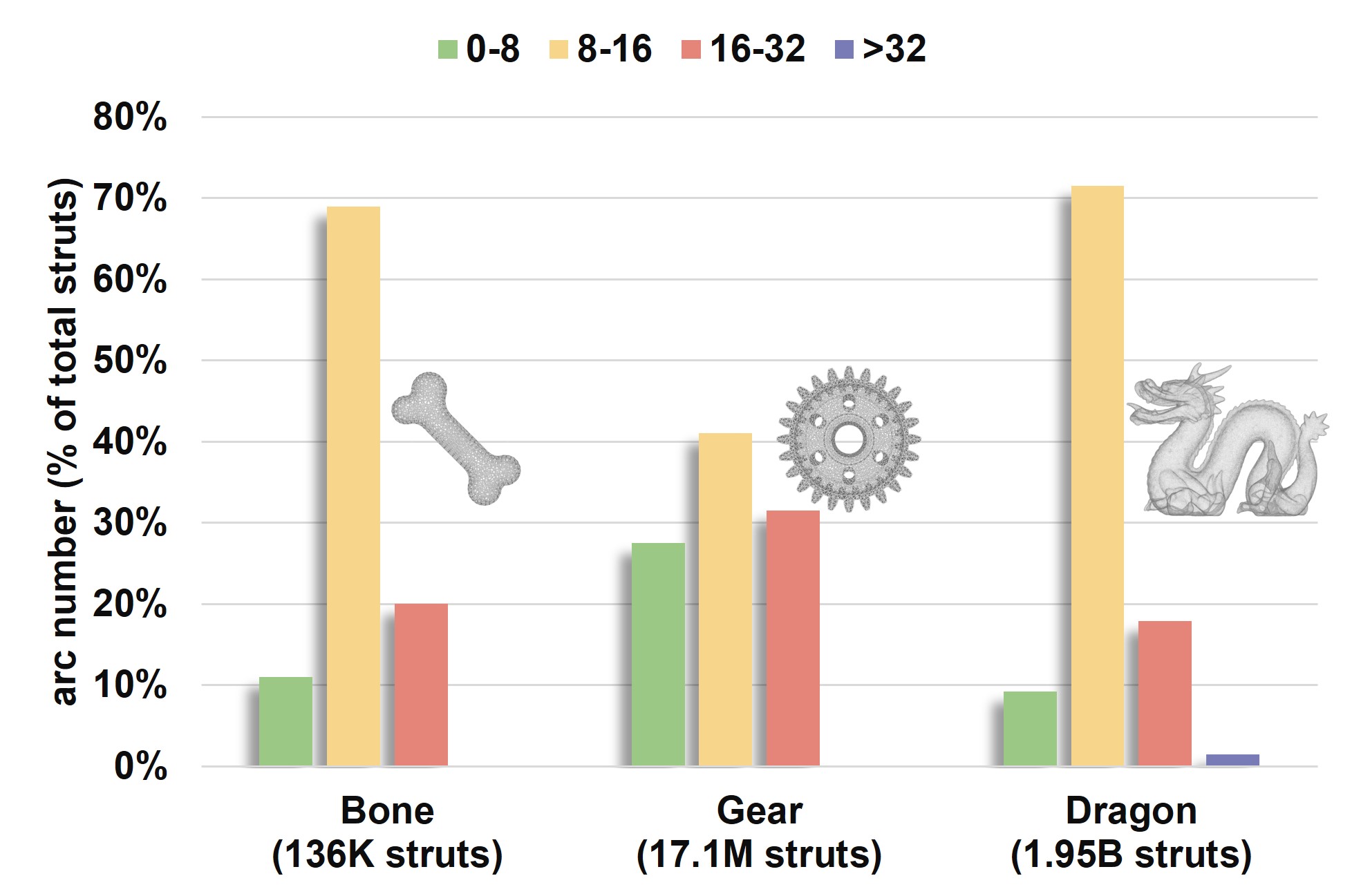}
	\caption{\rev{Statistics of the arc loop length on lattice struts.}{Statistics of the number of arcs of the bone, gear and dragon lattice structures from Fig.~\ref{fig: shell}.}}
 \label{fig: workload}
\end{figure}

\begin{figure*}[t]
\centering
\subfigure[]{
    \includegraphics[width = 0.32\textwidth]{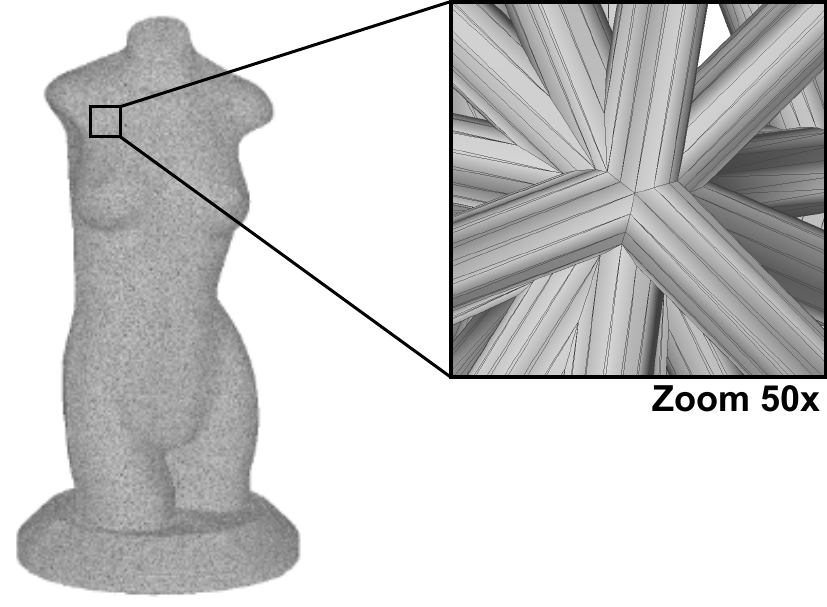}
    \label{fig: venus}
}
\subfigure[]{
    \includegraphics[width = 0.32\textwidth]{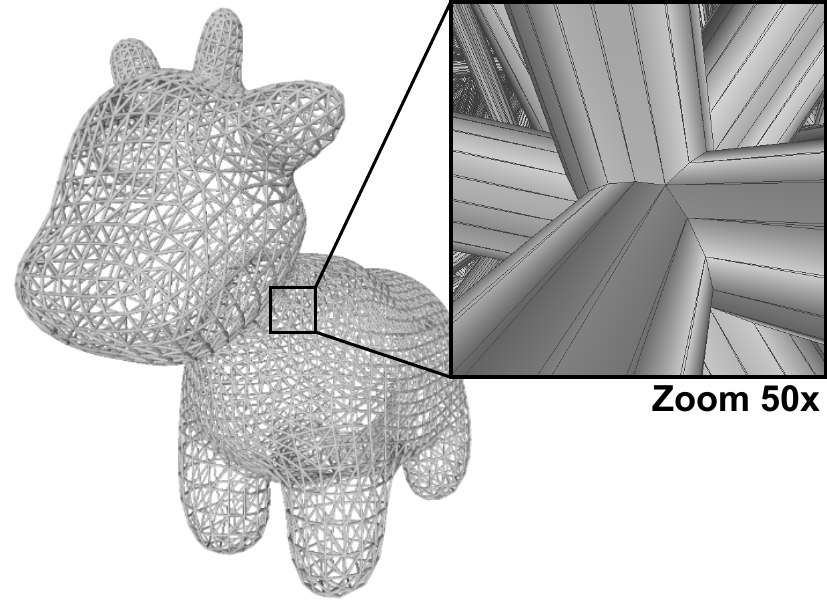}
    \label{fig: cow}
}
\subfigure[]{
    \includegraphics[width = 0.32\textwidth]{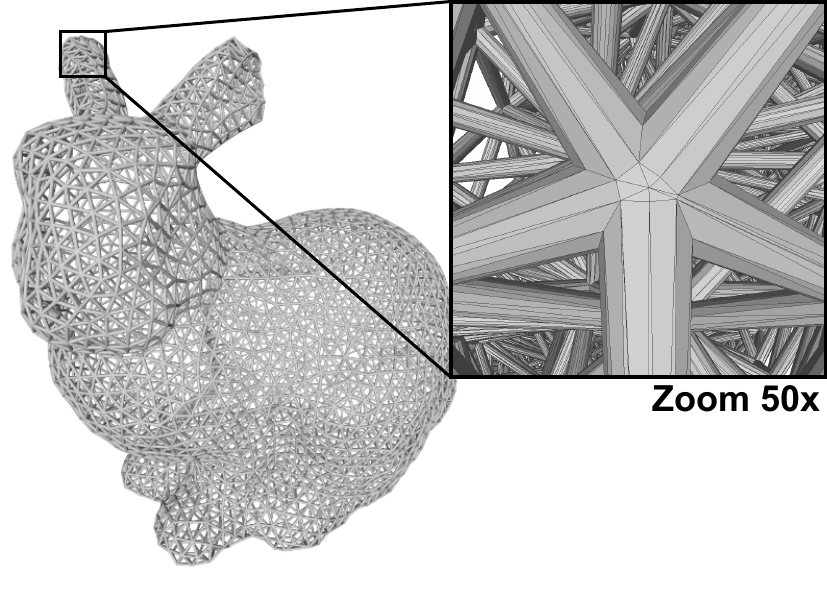}
    \label{fig: bunny}
}
\subfigure[]{
    \includegraphics[width = 0.32\textwidth]{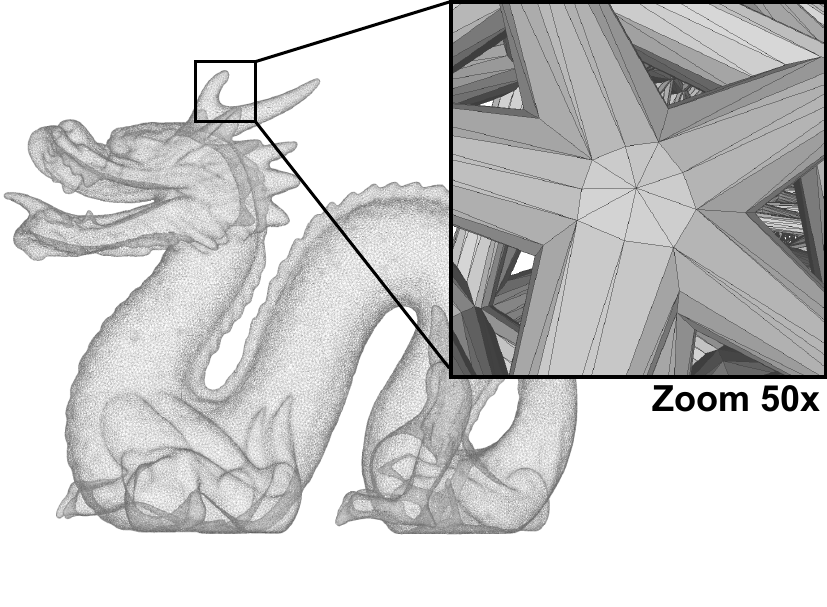}
    \label{fig: dragon}
}
\subfigure[]{
    \includegraphics[width = 0.32\textwidth]{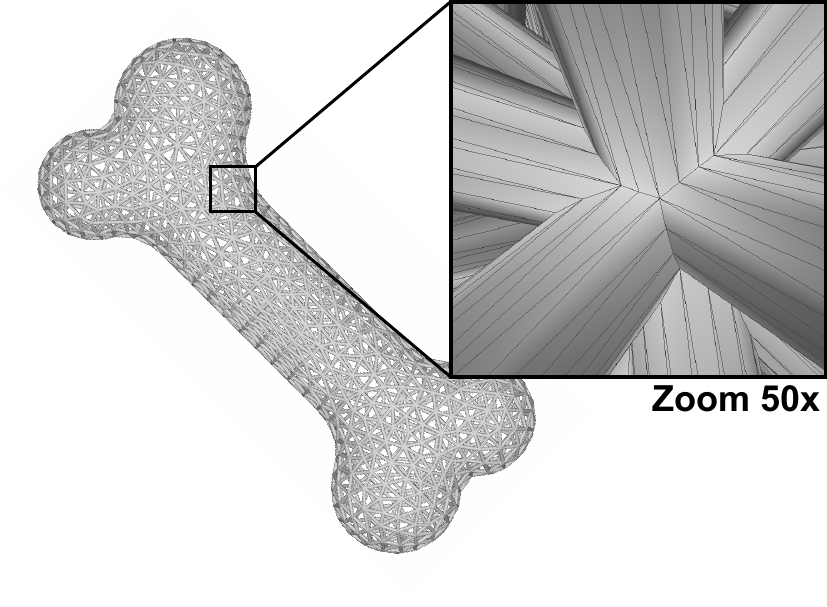}
    \label{fig: bone}
}
\subfigure[]{
    \includegraphics[width = 0.32\textwidth]{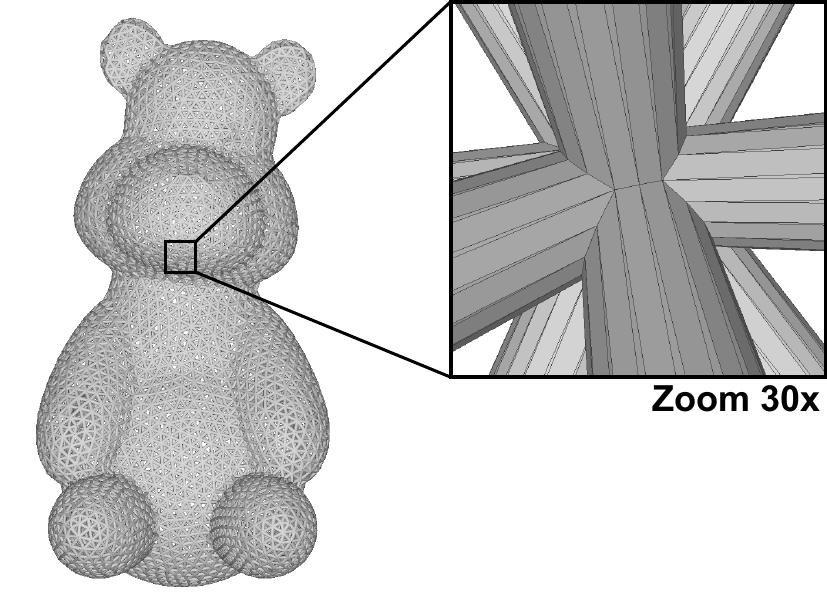}
    \label{fig: teddy}
}
\subfigure[]{
    \includegraphics[width = 0.32\textwidth]{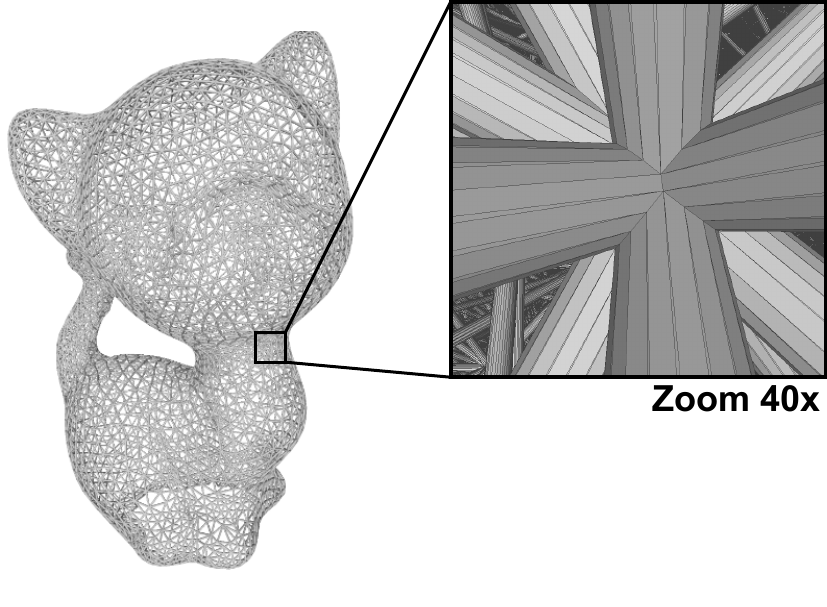}
    \label{fig: kitten}
}
\subfigure[]{
    \includegraphics[width = 0.32\textwidth]{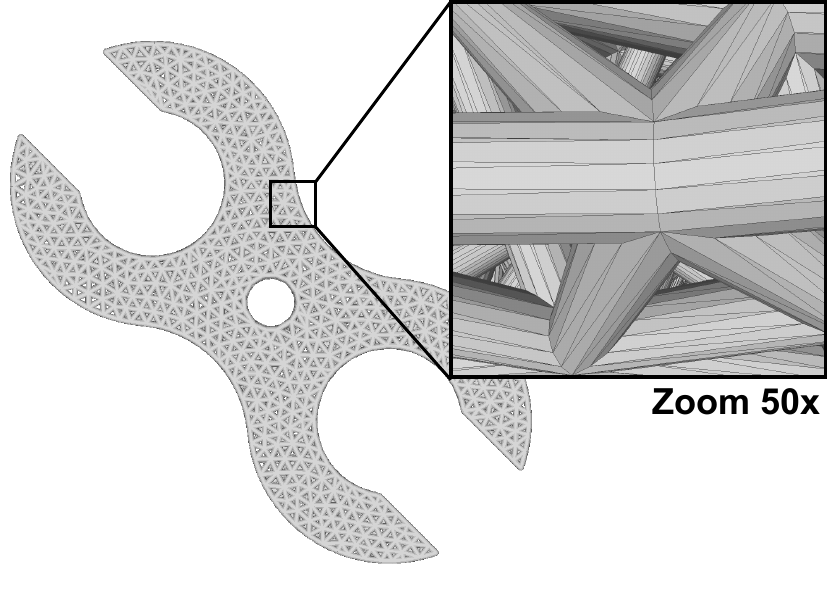}
    \label{fig: connector}
}
\subfigure[]{
    \includegraphics[width = 0.32\textwidth]{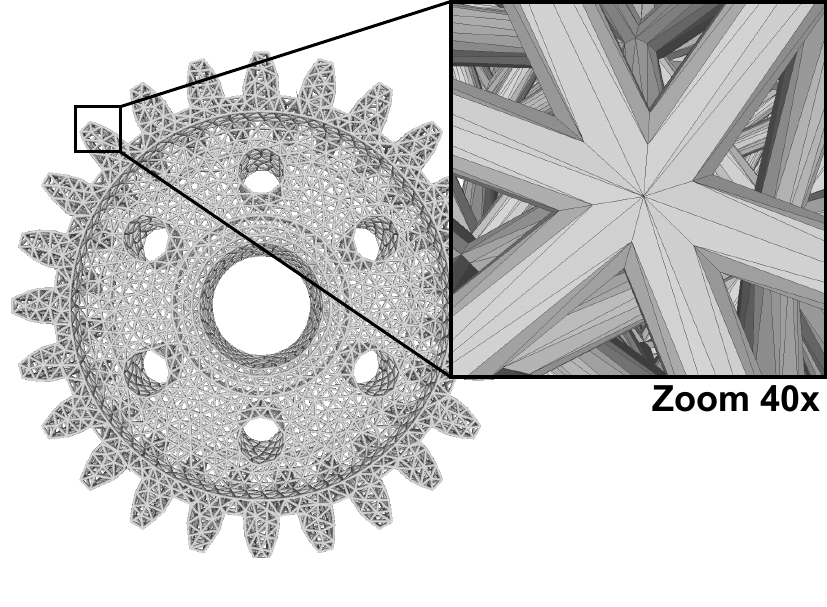}
    \label{fig: gear}
}
\caption{Models for case studies: (a) Venus with 0.5 billion struts; (b) Cow with 1 billion struts; (c) 
Stanford bunny with 1.4 billion struts; (d) Dragon with 1.9 billion struts; (e) Bone with 0.13 million struts; (f) Teddy bear with 1.5 million struts; (g) Kitten with 17 million struts; (h) Toy-connector with 3 million struts; (I) Gear with 17 million struts. Note that the rendered pictures here were based on simplified versions of billion-scale models for the sake of visual clarity; the actual billion-scale models are shown in the zoomed-in pictures.}
\label{fig: shell}
\end{figure*}

Lattice structures have complex connectivity, and the number of adjacent struts that connect to any common strut is varied, leading to varied quantities of strut-plane intersection for struts (or simply called workloads). \rev{From the statistics in Fig.~\ref{fig: workload}, more than 70\% of them are lower than 16.}{We collected struts from the bone, gear and dragon models in Fig.~\ref{fig: shell}, and the statistics (Fig.~\ref{fig: workload}) show that more than 70\% of them are lower than 16 in the arc number.}  However, the warp size of a typical GPU is 32, using one warp to process one strut would make nearly half of the threads in a warp inactive, resulting in low parallelism.

To maximize warp utilization and gain massive parallelism, warp-centric scheduling is utilized for the meta-meshing of the lattice structure. It combines the meta-meshing tasks of multiple struts and assigns them to one warp, making as many threads active as possible. For example, if two combined struts both have workloads of 16, then a warp can be fully used to handle the two struts. To achieve this objective, a sorting and scanning solution is proposed. Sorting facilitates the positioning of combinable struts adjacent to each other, and then scanning assigns struts that meet the combinable requirement to warps. \rev{}{The specific sorting and scanning algorithms used are Radix sorting and Blelloch scanning, which are both established and efficient algorithms~\cite{2024_CUDA-programming,2019_han_learn}.}

There still exists some struts that have over 32 workloads, as shown in Fig.~\ref{fig: workload}. Using a warp to process such struts demands a considerable amount of time. This results in temporary underutilization of warps, where completed warps must await uncompleted ones. For such a case, the scheduler employs multiple warps to meta-mesh a single strut, and then synchronizes these warps through shared memory. This scheduling approach enables efficient meta-meshing of struts with more than 32 curves, thereby mitigating temporary warp underutilization.

\section{Triangulation}
\label{sec:triangulation} 
\begin{figure}[b]
	\centering
	\includegraphics[width=0.4\textwidth]{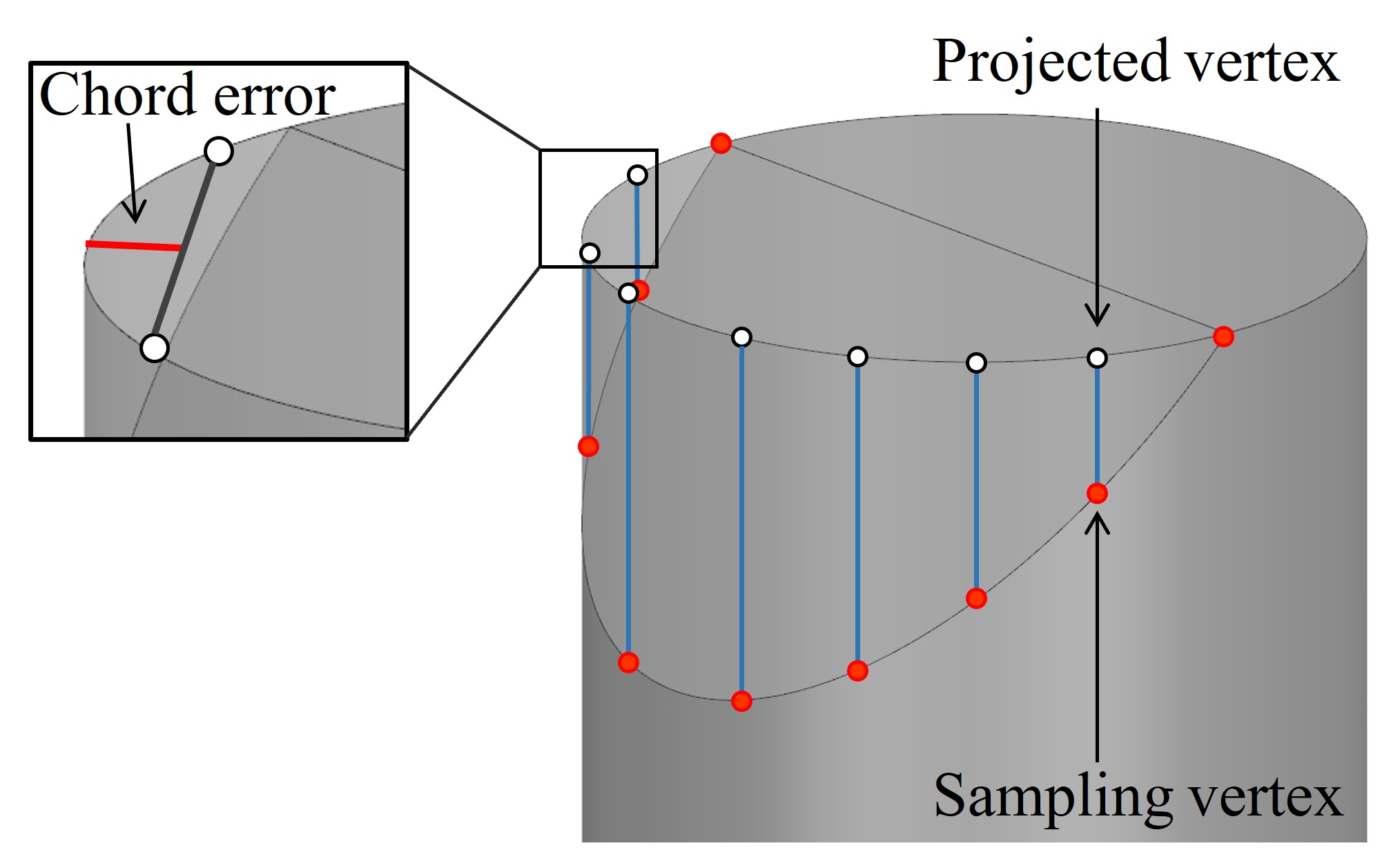}
	\caption{Illustration of subdividing arcs according to the chord error.}
 \label{fig: sampling}
\end{figure}

Based on the generated meta-mesh, triangulating lattice structures can be simply done by subdividing the meta-mesh's arcs. To make subdivided vertices uniformly distributed on the arc and therefore obtain a high-quality triangulation, the arc is projected to the circle edge of the strut's end section, and the subdivision process is implemented on the circle edge with controllable chord error. Then these subdivided vertices are mapped to the original arcs, as shown in Fig.~\ref{fig: sampling}. \rev{}{If we want to reuse the meta-mesh to generate a triangulation at a different resolution, we only need to re-subdivide the arcs according to the new chord error.}

Assuming the chord error (measured in terms of the percentage of the radius) is $CE$, and the parameter range of the arc is from $t_1$ to $t_2$, the number of the subdivided vertices $N$ is given by:
\begin{equation}
\label{equ: sampling}
\begin{split}
    N = \left\lfloor \cfrac{t_2-t_1}{2cos^{-1}(1-CE)} \right\rfloor + 1 \\
\end{split}
\end{equation}
where $\lfloor\rfloor$ is a floor function to get the integer part of the value. The formula finds the minimum number of vertices, which satisfy the specified chord error.
In practical applications, multiple triangulation results are required, like visualization. 
With uniform subdivision, we can define the parameters \rev{$t_i$}{$s_i$} on the original arcs by:
\begin{equation}
\label{equ: parameter-range}
\begin{split}
    s_i = \cfrac{t_2-t_1}{N}\times i + t_1 \quad \quad i = 0,1,...,N \\
\end{split}
\end{equation}

Using the above equation, all arcs on the two arc loops of a strut can be subdivided. Then a strut can be easily triangulated by connecting these vertices. Since the meta-mesh of each strut is independent, multiple CPU cores can be leveraged to accelerate this process significantly. The time of the triangulation overlaps with the meta-meshing in GPU and is fully hidden in the pipeline, with no more extra idle time needed.

\begin{figure}[t]
	\centering
    \subfigure[]{
        \includegraphics[width = 0.22\textwidth]{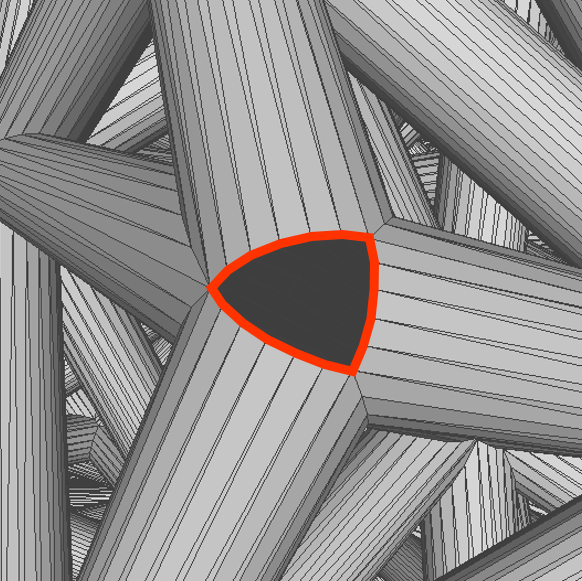}
        \label{fig: no-fill-hole}
    }
    \subfigure[]{
        \includegraphics[width = 0.22\textwidth]{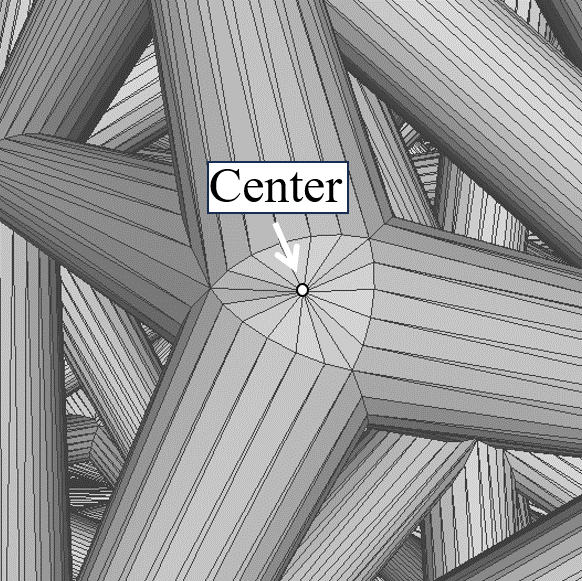}
        \label{fig: fill-hole}
    }
    \caption{Hole filling: (a) extracting the hole contour; (b) filling the hole by connecting the center and boundary vertices.}
    \label{fig: hole}
\end{figure}

While triangulation is completed, there may still exist holes among adjacent struts, as shown in Fig.~\ref{fig: hole}a.
\rev{}{The indication of holes lies in the special circular arcs that are at the end of struts and have no intersection with other struts, as shown in Figs.~\ref{fig: meta-mesh}b and \ref{fig: ellipse}. Therefore, we can simply add a one-bit flag to label such circular arcs, which enables automatic identification of holes in the triangulation stage.} \rev{This paper takes a straightforward method to fill these holes}{With holes identified, we then take a straightforward method to fill them}: (1) computing a barycenter of the vertices on the hole contour; (2) projecting the barycenter to its nodal sphere; (3) connecting the contour vertices to the projected center.
\rev{Fig.~\ref{fig: hole}b shows the results after hole filling.}{Assume a hole contour is formed by vertices $\boldsymbol{v}_0,\boldsymbol{v}_1,...,\boldsymbol{v}_{M-1}$, the barycenter $\boldsymbol{b}$ and the projected center $\boldsymbol{b}_{project}$ are given by:} 
\begin{equation}
\begin{split}
    \boldsymbol{b} &= \cfrac{\sum_{i=0}^{M-1}\boldsymbol{v}_i}{M} \\
    \boldsymbol{b}_{project} &= \cfrac{\boldsymbol{b} - \boldsymbol{o}}{||\boldsymbol{b}-\boldsymbol{o}||}\times r + \boldsymbol{o}\\
\end{split}
\end{equation}
\rev{}{where $\bm{o}$ is the center of the nodal sphere with radius $r$. After connecting vertices to the projected center, the hole is filled as shown in Fig.~\ref{fig: hole}b. It should be noted here that, this hole-filling method is chosen for no particular reason but a feasible one. Our primary contribution lies in the meta-meshing algorithm and pipeline presented in Sec.~\ref{sec:methods}. If there are alternative hole-filling methods, they can be readily integrated into the pipeline without affecting the pipe's effectiveness.}

\rev{}{
Based on the steps mentioned above, the triangulation process is implemented using Algorithm~\ref{algo: triangulation}. The algorithm accepts the meta-mesh and a list of chord errors as the input, and outputs triangulation results with desired resolutions. Starting from specified error $CE_i$, the first step subdivides arcs into vertices~(Line 5) and records vertices on the hole contour~(Line 6). Then, struts are triangulated by connecting subdivided vertices~(Line 7). Finally, it fills holes and writes the result triangles to the disk~(Lines 9-10).}
\begin{algorithm}[htbp]
        \caption{Triangulation}
        \label{algo: triangulation}
        \begin{algorithmic}[1] 
        \Require meta-mesh $MM$ and chord error list $CL = \{CE_{1}, CE_{2}, \cdots, CE_{n}\}$
        \Ensure lattice structure triangulation results
        \For{each chord error $CE_{i}$ in $CL$}
            \State $T \leftarrow \emptyset$
            \State $Holes \leftarrow \emptyset$
            \For{each arcloop $AL$ in $MM$}
                \State $VL \leftarrow $ subdivide($AL, CE_{i}$) 
                \State $Holes \leftarrow Holes$ $\cup$ holesDetect($VL$)
                \State $T \leftarrow T$ $\cup$ triangulate($VL$)
            \EndFor
            \State $T \leftarrow T \cup$ fillHoles($Holes$)
            \State write($T$) \Comment{Output results to a file}
        \EndFor
    \State \Return{}
    \end{algorithmic}
\end{algorithm}

\def\tabularxcolumn#1{m{#1}}
\begin{table*}[htb]
\caption{Statistics of models generated by the meta-meshing method.}
\centering
\setlength\extrarowheight{5pt}
\begin{tabularx}{1.0\textwidth}{
    >{\arraybackslash}>{\hsize=.45\hsize\linewidth=\hsize}X
    >{\centering\arraybackslash}>{\hsize=.3\hsize\linewidth=\hsize}X
    >{\centering\arraybackslash}>{\hsize=.3\hsize\linewidth=\hsize}X
    >{\centering\arraybackslash}>{\hsize=.3\hsize\linewidth=\hsize}X
    >{\centering\arraybackslash}>{\hsize=.3\hsize\linewidth=\hsize}X
    >{\centering\arraybackslash}>{\hsize=.3\hsize\linewidth=\hsize}X
    >{\centering\arraybackslash}>{\hsize=.3\hsize\linewidth=\hsize}X
    >{\centering\arraybackslash}>{\hsize=.3\hsize\linewidth=\hsize}X
    >{\centering\arraybackslash}>{\hsize=.3\hsize\linewidth=\hsize}X
    >{\centering\arraybackslash}>{\hsize=.3\hsize\linewidth=\hsize}X
   }
    \Xhline{1pt}
    Model &Venus &Cow &Bunny &Dragon &Bone &Teddy &Kitten &Connector &Gear\\
    \Xhline{1pt}
    Strut         &0.51B  &1.02B &1.40B   &1.95B  &136K   &1.57M &17.3M &3.19M  &17.1M \\
    Avg. strut len&0.3mm  &0.05mm &0.05mm &0.05mm &1.0mm  &1.0mm &1.0mm &0.1mm  &0.1mm \\
    Radius        &0.05mm &0.01mm &0.01mm &0.01mm &0.2mm  &0.2mm &0.2mm &0.02mm &0.02mm\\
    Chord error   &5\%    &3\%    &5\%    &5\%    &2\%    &2\%   &2\%   &6\%    &6\%   \\
    File size     &735G   &1,183G &1,617G &2,376G &241M   &2.72G &29.5G &3.54G  &16.7G \\
    Time          &617s   &995s   &1,091s &1,868s &0.118s &1.20s &15.7s &1.93s  &9.36s \\
    \Xhline{1pt}
\end{tabularx}
\label{tab: models}
\end{table*}

\begin{figure*}[htbp]
	\centering
    \subfigure[]{
        \includegraphics[width = 0.32\textwidth]{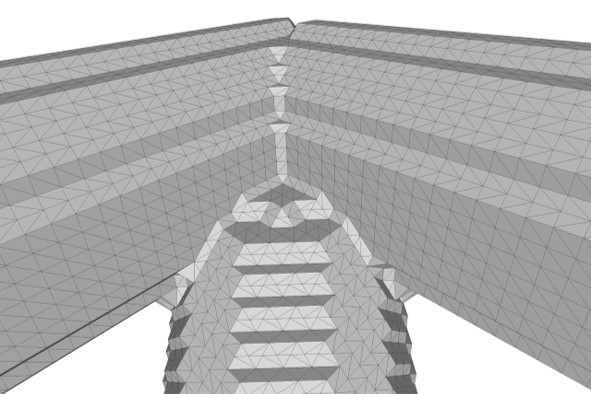}
        \label{fig: detail_MCM}
    }
    \subfigure[]{
        \includegraphics[width = 0.32\textwidth]{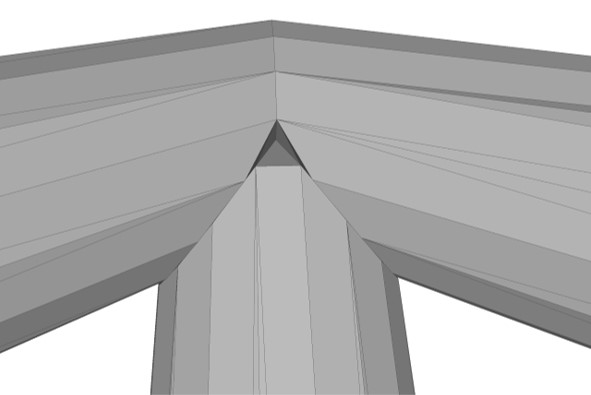}
        \label{fig: detail_LSLT}
    }
    \subfigure[]{
        \includegraphics[width = 0.32\textwidth]{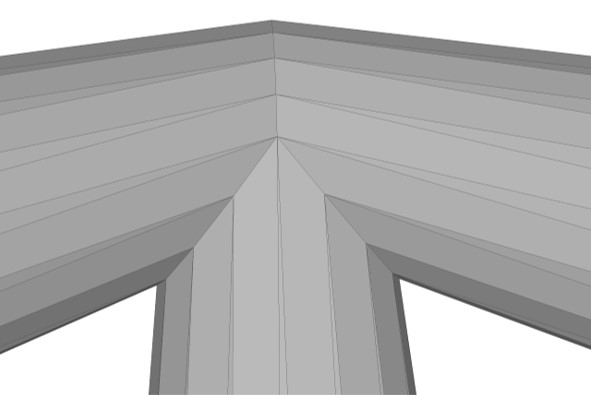}
        \label{fig: detail_meta-mesh}
    }
    \caption{\rev{}{Comparisons of triangulation quality between MC, LSLT, and meta-meshing methods using the Bone model: (a) MC method; (b) LSLT method; (c) meta-meshing method.}}
    \label{fig: vs-detail}
\end{figure*}

\begin{table*}[htb]
\caption{Comparisons of MC, LSLT, and meta-meshing method.}
\centering
\setlength\extrarowheight{8pt}
\begin{tabularx}{1.0\textwidth}{
    >{\centering\arraybackslash}>{\hsize=.5\hsize\linewidth=\hsize}X
    >{\centering\arraybackslash}>{\hsize=.5\hsize\linewidth=\hsize}X
    >{\centering\arraybackslash}>{\hsize=.5\hsize\linewidth=\hsize}X
    >{\centering\arraybackslash}>{\hsize=.5\hsize\linewidth=\hsize}X
    >{\centering\arraybackslash}>{\hsize=.5\hsize\linewidth=\hsize}X
    >{\centering\arraybackslash}>{\hsize=.5\hsize\linewidth=\hsize}X
    >{\centering\arraybackslash}>{\hsize=.5\hsize\linewidth=\hsize}X
    >{\centering\arraybackslash}>{\hsize=.5\hsize\linewidth=\hsize}X
   }
    \Xhline{1pt}
    Model &Strut &\makecell[c]{Triangles\\(MC)} &\makecell[c]{Triangles\\(LSLT)} &\makecell[c]{Triangles\\(Meta-mesh)} &\makecell[c]{Time\\(MC)} &\makecell[c]{Time\\(LSLT)} &\makecell[c]{Time\\(Meta-mesh)}\\
    \Xhline{1pt}
    Bone   &136,391    &262.4M  &8.37M  &4.78M &11.2s  &19.9s  &0.118s\\
    Teddy  &1,568,301  &1,222M  &97.4M  &54.8M &101.8s &231s   &1.20s\\
    Kitten &17,331,861 &Crashed &1,092M &595M  &N/A    &2,461s &15.7s\\
    \Xhline{1pt}
\end{tabularx}
\label{tab: LSLT-vs-Meta-mesh}%
\end{table*}

\section{Results and Discussion}\label{sec: results}
To test the performance of the proposed method, a Nvidia RTX 3090 GPU paired with an Intel Core i9-12900K CPU running at 5.20GHz is chosen. The GPU is equipped with 82 streaming multiprocessors (SMs), 128 KB of L1 cache, 6 MB of L2 cache, and a total of 24 GB of global memory. The method has been implemented using C++ 17 and CUDA 12.2, operating within the Ubuntu 20.04 environment. 

Based on this implementation, 9 case studies are to be presented to demonstrate the effectiveness of the proposed method. Case studies 1-4 (Fig.~\ref{fig: shell}a-d) considered four large lattice structures that have serval billion struts. Case studies 5-7 (Fig.~\ref{fig: shell}e-g) considered three small-scale lattice structures, ranging from 0.1 to 10 million. They were used to carry out comparisons with the state-of-the-art method~\cite {2017_Chougrani_lightweight-triangulation}. The last two case studies 8-9 (Fig.~\ref{fig: shell}h-i) used two real-world machine components. These two models were used to verify whether the challenges were resolved.
All these models were generated by meshing boundary models using tetgen~\cite{2015_Hang_Tetgen}. For billion-scale models, which are too large to be tetrahedralized, we generated million-scale models and then subdivided them to achieve the billion-scale counterparts.

\subsection{Examples}
\label{sec: examples}
Case study 1 (Fig.~\ref{fig: shell}a) is a Venus model with bounding box of 160mm $\times$ 80mm $\times$ 80mm. Case study 2 (Fig.~\ref{fig: shell}b) considered a cow model with a total of one billion struts. This model has a bounding box of 90mm $\times$ 150mm $\times$ 150mm. Case study 3 (Fig.~\ref{fig: shell}c) analyzed a bunny model with 1.4 billion struts. The size of the case is set to 120mm $\times$ 150mm $\times$ 150mm. Case study 4 (Fig.~\ref{fig: shell}d) is more complex, involving 2 billion struts and a bounding box of 90mm $\times$ 140mm $\times$ 200mm. They are used to evaluate the performance of billion-scale lattice structures. Moreover, these four models have different strut radii and approximation errors. 

Case study 5 (Fig.~\ref{fig: shell}f) considered a bone model with 0.13 million struts. Case study 6 (Fig.~\ref{fig: shell}g) analyzed a teddy model, involving 1.5 million struts, and case study 7 (Fig.~\ref{fig: shell}h) is a kitten model characterized by 17.3 million struts. They all have a strut radius of 0.2mm and chord error of 2\%. \rev{The three models are used for comparison with the LSLT method~\cite{2017_Chougrani_lightweight-triangulation}.}{The three models are used for comparing with the marching cubes~(MC) and LSLT method~\cite{2017_Chougrani_lightweight-triangulation}.} We counted the triangulation runtime and the number of triangles generated as summarized in Table~\ref{tab: LSLT-vs-Meta-mesh}. 
\rev{}{The triangulation quality of the three methods has also been compared, as shown in Fig.~\ref{fig: vs-detail}.}
The proposed method is seen to provide a significant improvement to the state of the art.

\begin{figure}[htb]
	\centering
	\includegraphics[width=0.42\textwidth]{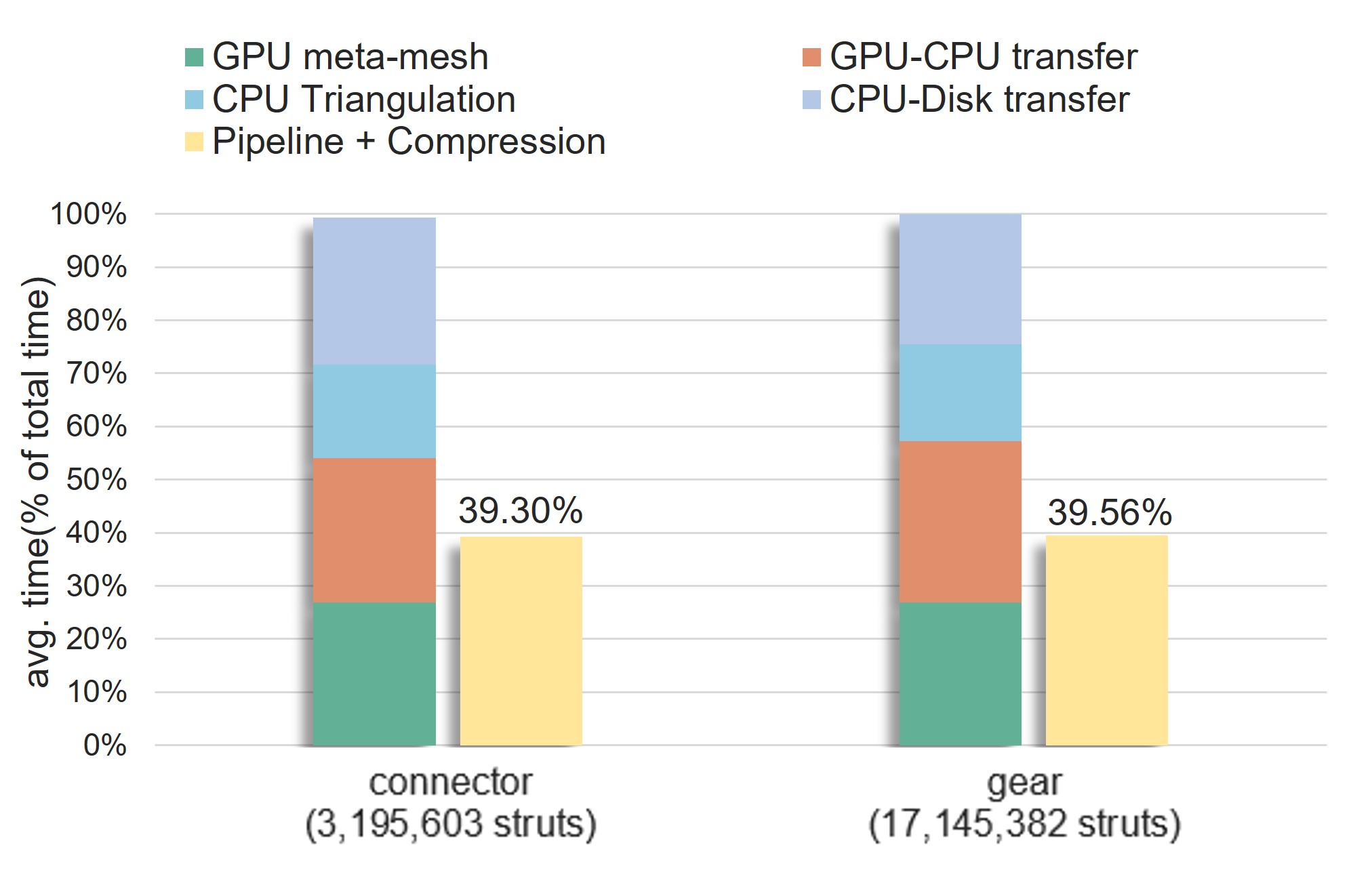}
	\caption{Comparisons between the non-pipeline implementation and the asynchronous pipeline with data compression.}\label{fig: pipeline-vs-non}
\end{figure}

\begin{figure}[htb]
	\centering
	\includegraphics[width=0.415\textwidth]{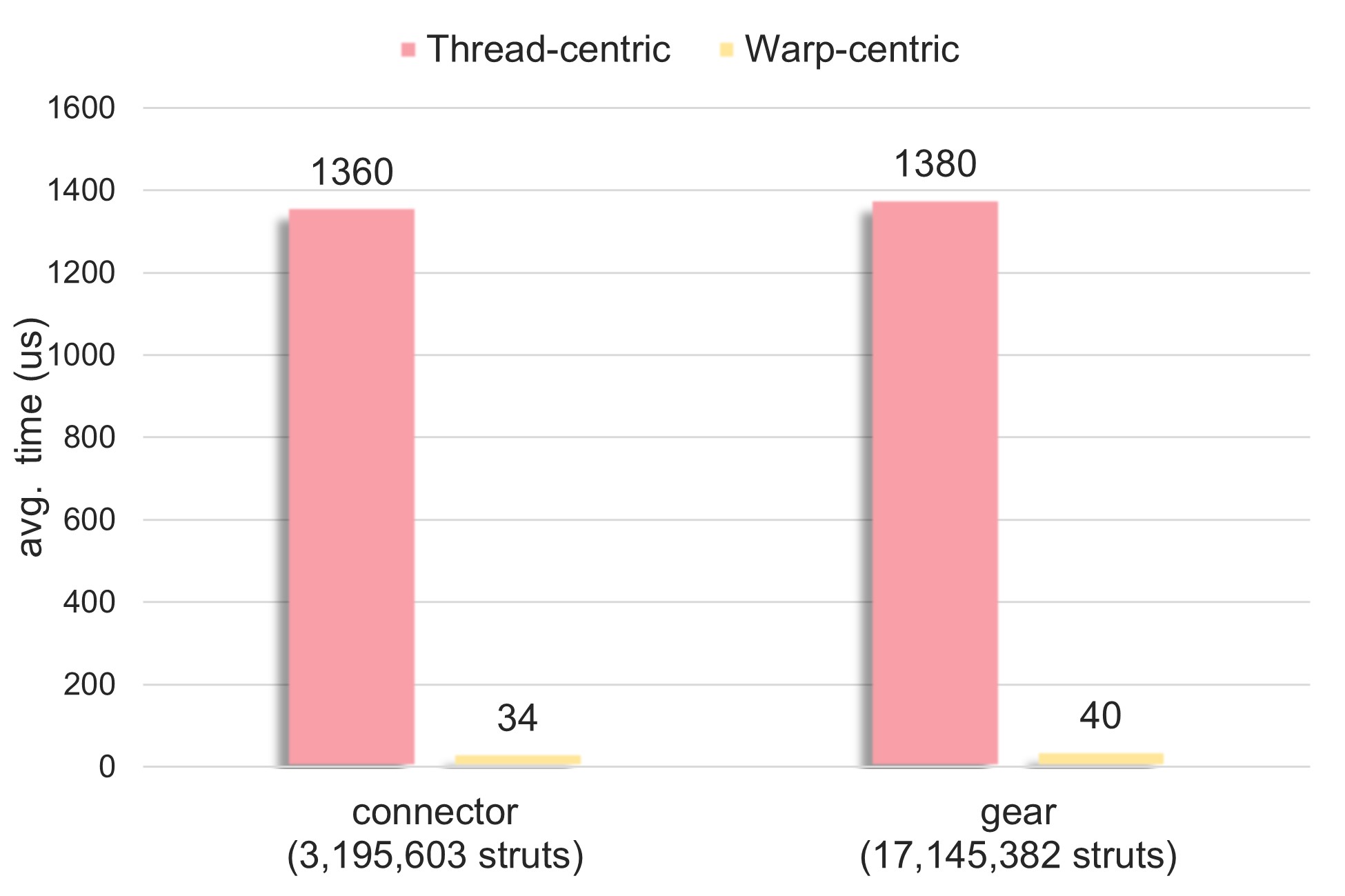}
	\caption{Comparisons between thread-centric and warp-centric meta-meshing.}\label{fig: thread-vs-warp}
\end{figure}

Case studies 8 (Fig.~\ref{fig: shell}h) and 9 (Fig.~\ref{fig: shell}i) considered two real-world machine parts. Case study 8 is a toy connector with 3.2 million struts and a bounding box of 16mm $\times$ 16mm $\times$ 32mm. Case study 9 analyzed a gear model, involving 17.1 million struts, with a size of 80mm $\times$ 80mm $\times$ 20mm. The radius (0.02mm) and chord error (6\% of radius) are chosen for both models. Both models can be downloaded from the GrabCAD library (https://grabcad.com/library). To see whether the meta-meshing method resolves data transfer and computational challenges mentioned in Sec.~\ref{sec:intro}, we conducted ablation experiments on case studies 8-9. The runtime of each pipeline stage was collected to verify whether the transfer bottleneck had been overcome. For GPU-Lattices structural mismatch, the traditional thread-centric method was used for comparison with our meta-meshing method. The comparison results are shown in Figs.~\ref{fig: pipeline-vs-non} and~\ref{fig: thread-vs-warp}.

\rev{}{In the above comparisons, the thread-centric meta-meshing scheme was implemented by following the traditional thread-based algorithm design paradigm. It takes a weighted graph representation of the lattice structure as input and then allocates a thread for an arc calculation. During arc calculation, every thread accesses the relevant strut parameters stored in the global memory by itself, without considering other threads’ running status. All scheduling and memory access optimization tasks are left to CUDA, if any. The instructions used to calculate the specific intersection arc for two given struts are the same as those used in our warp-centric scheme, so is the data transfer between the CPU and the GPU.}

\par

\subsection{Discussions and limitations}
\begin{figure*}[htb]
	\centering
	\includegraphics[width=0.95\textwidth]{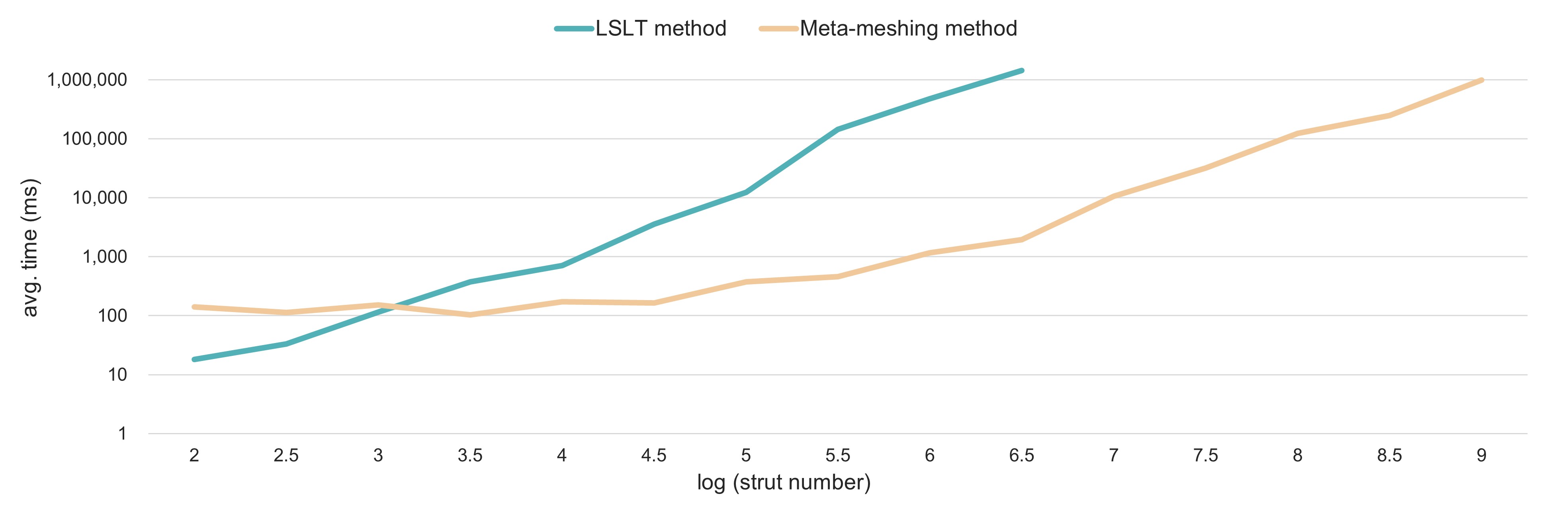}
	\caption{Comparisons of computational efficiency between meta-meshing method and LSLT method as the number of struts increases.}
 \label{fig: limitation}
\end{figure*}
For all examples, the proposed method can efficiently and robustly generate meta-meshes and triangular meshes. 
This method completes the triangulation of the cow model with one billion struts within 17 minutes and exhibits excellent scalability in different models. Even when processing the most complex dragon model with 2 billion struts, triangulation can be completed within 30 minutes (Table~\ref{tab: models} and Fig.~\ref{fig: limitation}). 
By solving both data transfer and computational bottlenecks on GPU, meta-meshing becomes even more efficient. Through a CPU-GPU asynchronous pipeline and compression, the entire pipeline achieved a 2.5x speedup (Fig.~\ref{fig: pipeline-vs-non}). Compared with the traditional thread-centric paradigm, the GPU kernel execution in a warp-centric way is 35x faster than the thread-centric way (Fig.~\ref{fig: thread-vs-warp}).

\rev{In addition to its efficiency, the proposed method produces lightweight outcomes. In comparison with the LSLT method, our analysis reveals that the meta-meshing method generates much fewer triangle facets under the same approximation error. This outcome can be attributed to the meta-mesh’s precise representation of struts within the lattices, leading to a minimized number of sampling vertices. As the number of sampled vertices decreases, so does the number of triangulated facets. As detailed in Table~\ref{tab: LSLT-vs-Meta-mesh}, when both the meta-meshing method and the LSLT method utilize a 2\% chord error, the number of triangles generated by the meta-meshing method amounts to approximately 50\% of that generated by the LSLT method. Notably, this ratio is expected to further decrease with a larger chord error.}
{In addition to high efficiency, the proposed method produces lightweight triangular meshes. Compared with MC and LSLT, our method generates fewer triangles under the same approximation error~(Table~\ref{tab: LSLT-vs-Meta-mesh}). This is attributed to the meta-mesh’s precise representation of struts. As can be seen from Fig.~\ref{fig: vs-detail}, the MC method's mesh contains many redundant triangle facets and is fairly blocky. The LSLT method's mesh does not align vertices between struts, resulting in additional triangle faces. In the case of using a 2\% chord error, the number of triangles generated by the meta-meshing method is approximately 95\% fewer than the MC method and about 50\% fewer than the LSLT method. This ratio is expected to decrease further if a larger chord error is used.} 

\rev{}{Some remarks should be noted here. For the LSLT method, it would be much better to have a GPU version of it and then compare this version with the proposed method. Unfortunately, there are two issues hindering the development of the GPU version.  First, we have done an extensive search on the internet but found no publicly available codes for LSLT---the method’s authors have not released their codes yet. Second, the original LSLT method was designated for serial computing, and consequently converting its intrinsic serial logic into a parallel logic is not straightforward; thus further development is required. In these regards, a comparison with the GPU version of LSLT were not included in the present work.}

Despite the excellent performance of the meta-meshing method on large-scale lattice structures, the entire pipeline often incurs startup time. For small-scale data, using CPU processing alone proves more efficient and straightforward. Fig.~\ref{fig: limitation} illustrates the comparison between the meta-mesh and LSLT methods regarding the processing time for lattice structures from hundred to billion struts. It can be observed that the meta-meshing method requires a startup time of 100ms-200ms in our hardware configuration. Therefore, in cases where the number of struts is less than thousands, a CPU method may be easier and more direct to triangulate the lattice structure. On the other hand, when the number of struts reaches 1 million or even 1 billion, the meta-meshing method has a much better performance than traditional CPU triangulation methods.

\section{Conclusion}
\label{sec:conclusion}
This paper presents a new method called meta-meshing for solving the billion-scale lattice structure triangulation problem. The main feature of this method is its high efficiency, which can complete the triangulation of lattice structures with a billion struts in minutes. 
This feature is essentially achieved by using the meta-mesh representation, which is lightweight and can conveniently triangulate lattice structures with arbitrary resolutions. By resolving challenges in GPU-CPU data transfers, lattice-GPU mismatch, and warp divergence, a high-performance CPU-GPU meta-meshing pipeline is developed. A notable improvement over existing methods has been demonstrated in various experiments.

It should be noted that the proposed method adopts a CPU+GPU collaborative architecture. This architecture cooperatively utilizes GPU, CPU, and disk to perform a task. However, the poor performance of a single device can impact the efficiency of the entire pipeline. This is thus a serious limitation of the proposed method. Moreover, the current meta-meshing method focuses solely on the lattice structures with cylindrical and conical struts. There are still other types of lattice structures that do not use cylinders and cones as struts, such as Quador~\cite{2018_Gupta_Quador}. Extending the proposed method to handle more complex lattice structures is an interesting improvement direction in the future. 

Another interesting improvement is to combine meta-meshing with the very recent direct CAD modeling paradigm~\cite{zou2019push,zou2020decision,zou2019variational,zou2022robust}. As such, the user can intuitively and freely edit a portion of a lattice structure, and then the computer can promptly update its appearance, locally and adaptively. Without meta-meshing, interaction-demanding applications such as direct modeling are impossible.

\section*{Acknowledgements}
This work has been funded by NSF of China (No. 62102355), the ``Pioneer" and ``Leading Goose" R\&D Program of Zhejiang Province (No. 2024C01103), NSF of Zhejiang Province (No. LQ22F020012), and the Fundamental Research Funds for the Zhejiang Provincial Universities (No. 2023QZJH32).


\section*{References}

\bibliographystyle{elsarticle-num}
\bibliography{paperRef}




\end{document}